\documentclass[twocolumn,showpacs,preprintnumbers,amsmath,amssymb]{revtex4}

\usepackage{graphicx}
\usepackage{dcolumn}
\usepackage{bm}

\begin{document}

\title{Meissner phases in spin-triplet ferromagnetic superconductors}

\author{Diana V. Shopova}
 \altaffiliation[]{Corresponding author. Electronic address: sho@issp.bas.bg}
\affiliation{CP Laboratory, Institute of Solid State Physics, Bulgarian
Academy of Sciences, BG--1784 Sofia, Bulgaria}

\author{Dimo I. Uzunov}
 \altaffiliation[]{Electronic address: uzun@issp.bas.bg. Permanent address: CP Laboratory,
 Institute of Solid State Physics, Bulgarian Academy of Sciences, BG--1784, Sofia, Bulgaria.}
\affiliation{Max-Plank-Institut  f\"{u}r Physik komplexer Systeme,\\
N\"{o}tnitzer Str. 38, 01187 Dresden, Germany, and\\
Abdus Salam International Centre for Theoretical Physics,\\ Strada
Costiera 11, 34014 Trieste, Italy}

\date{25th March 2005}

\begin{abstract}
We present new results  for the properties of phases and phase
transitions in spin-triplet ferromagnetic superconductors. The
superconductivity of the mixed phase of coexistence of ferromagnetism
and unconventional superconductivity is triggered by the presence of
spontaneous magnetization. The mixed phase is stable but the other
superconducting phases that usually exist in unconventional
superconductors are either unstable or for particular values of the
parameters of the theory some of them are metastable at relatively low
temperatures in a quite narrow domain of the phase diagram. Phase
transitions from the normal phase to the phase of coexistence is of
first order while the phase transition from the ferromagnetic phase to
the coexistence phase can be either of first or second order depending
on the concrete substance. Cooper pair and crystal anisotropies
determine a more precise outline of the phase diagram shape and  reduce
the degeneration of  ground states of the system but they do not
 change drastically  phase stability domains and
thermodynamic properties of the respective phases. The results are
discussed in view of application to metallic ferromagnets as UGe$_2$,
ZrZn$_2$, URhGe.
\end{abstract}



\maketitle

\section{\label{sec:level1}Introduction}

In 2000, experiments~\cite{Saxena:2000} at low temperatures ($T \sim 1$
K) and high pressure ($P\sim 1$ GPa) demonstrated the existence of spin
triplet superconducting states in the metallic compound UGe$_2$. The
superconductivity is triggered by the spontaneous magnetization of the
ferromagnetic phase that occurs at much higher temperatures. It
coexists with the superconducting phase in the whole domain of its
existence below $T \sim 1$ K; see also experiments from
Refs.~\cite{Huxley:2001, Tateiwa:2001}, and the discussion in
Ref.~\cite{Coleman:2000}. The same phenomenon of existence of
superconductivity at low temperatures and high pressure in the domain
of the $(T,P)$ phase diagram where the ferromagnetic order is present
was observed in other ferromagnetic metallic compounds
(ZrZn$_2$~\cite{Pfleiderer:2001} and URhGe~\cite{Aoki:2001}) soon after
the discovery~\cite{Saxena:2000} of superconductivity in UGe$_2$.

In contrast to other superconducting materials, as ternary and Chevrel
compounds, where the influence of magnetic order on superconductivity
is also substantial (see, e.g.,~\cite{Vonsovsky:1982, Maple:1982,
Sinha:1984, Kotani:1984}), in these ferromagnetic substances the phase
transition temperature ($T_f$) to the ferromagnetic state is much
higher than the phase transition temperature ($T_{FS})$ from
ferromagnetic to a mixed state of coexistence of ferromagnetism and
superconductivity. For example, in UGe$_2$, $T_{FS} = 0.8$ K while the
critical temperature of the phase transition from paramagnetic to
ferromagnetic state in the same material is $T_f =35
$K~\cite{Saxena:2000, Huxley:2001}. It can be assumed that in these
substances
 the material parameter $T_s$ defined as the
usual critical temperature of the second order phase transition from
normal to uniform (Meissner) supercondicting state in a zero external
magnetic field is much lower than the phase transition temperature
$T_{FS}$. The above mentioned experiments on the compounds~UGe$_{2}$,
URhGe, and ZrZn$_2$ do not give any evidence for the existence of a
standard normal-to-superconducting phase transition in a zero external
magnetic field.

It seems that the superconductivity in the metallic compounds mentioned
above always coexists with the ferromagnetic order and is enhanced by
it. In these systems, as claimed in Ref.~\cite{Saxena:2000}, the
superconductivity probably arises from the same electrons that create
the band magnetism and can be most naturally understood rather as a
triplet than spin-singlet pairing phenomenon.  Metallic compounds
UGe$_{2}$, URhGe, and ZrZn$_2$, are itinerant ferromagnets. An
unconventional superconductivity is also suggested~\cite{Saxena:2001}
as a possible outcome of recent experiments in Fe~\cite{Shimizu:2001},
in which a superconducting phase has been  discovered at temperatures
below $2$ K and pressures between 15 and 30 GPa. There both vortex and
Meissner superconductivity phases~\cite{Shimizu:2001} are found in the
high-pressure crystal modification of Fe with a hexagonal close-packed
lattice for which the strong ferromagnetism of the usual bcc iron
crystal probably disappears~\cite{Saxena:2001}. It can be hardly
claimed that in hexagonal Fe the ferromagnetism and superconductivity
coexist  but the clear evidence for a superconductivity is also a
remarkable achievement.

The reasonable question whether these examples of superconductivity and
coexistence of superconductivity and ferromagnetism are bulk or surface
effects can be stated. The earlier experiments performed before 2004 do
not answer this question. Recent experiments~\cite{Yelland:2005} show
that surface superconductivity appears in ZrZn$_2$ and its presence
depends essentially on the way of preparation of the sample. But in our
study it is important that bulk superconductivity can be considered
well established in this substance.

A phenomenological theory that explains the coexistence of
ferromagnetism and unconventional spin-triplet superconductivity of
Landau-Ginzburg type has been developed recently in
~\cite{Machida:2001, Walker:2002} where possible low-order couplings
between the superconducting and ferromagnetic order parameters are
derived with the help of general symmetry group arguments. On this
basis several important features of the superconducting vortex state
of unconventional ferromagnetic superconductors were
established~\cite{Machida:2001, Walker:2002}.

In our paper we shall follow the approach from
Refs.~\cite{Machida:2001, Walker:2002}  to investigate the conditions
for the occurrence of the Meissner phase and to demonstrate that the
presence of ferromagnetic order enhances the $p$-wave
superconductivity. We also establish the phase diagram of
 ferromagnetic superconductors in a zero
external magnetic field and show that the phase transition to the
superconducting state  can be either of first or second order depending
on the particular substance. We confirm the predictions made in
Refs.~\cite{Machida:2001,Walker:2002} about the symmetry of the ordered
phases.

Our investigation is based on the mean-field
approximation~\cite{Uzunov:1993} as well as on known results about the
possible phases in nonmagnetic superconductors with triplet ($p$-wave)
pairing~\cite{Volovik:1985, Blagoeva:1990, Uzunov:1990, Sigrist:1991}.
We extend  our preceding results~\cite{Shopova1:2003, Shopova2:2003,
Shopova3:2003} and show that taking into account the anisotropy of the
spin-triplet Cooper pairs modifies but does not drastically change the
thermodynamic properties of the coexistence phase, especially in the
temperature domain above the superconducting critical temperature
$T_s$. The effect  of crystal anisotropy is similar but we shall not
make an overall thermodynamic analysis of this problem  because we have
to consider  concrete systems and crystal
structures~\cite{Volovik:1985, Sigrist:1991} for which there is no
enough information from experiment to make conclusions about the
parameters of the theory. Our results confirm the general concept that
the anisotropy reduces the degree of ground state degeneration, and
depending on the symmetry of the crystal, picks up a crystal direction
for the ordering.

There exists a formal similarity between the phase diagram we
obtain and the phase diagram of certain improper
ferroelectrics~\cite{Gufan:1980, Gufan:1981, Latush:1985,
Toledano:1987, Gufan:1987, Cowley:1980}. We shall make use of the
concept in  the theory of improper ferroelectrics, where the
trigger of the primary order parameter  by a secondary order
parameter (the electric polarization) has been initially
introduced and exploited; see Ref.~\cite{Toledano:1987,
Gufan:1987, Cowley:1980}. The mechanism of the M-triggered
superconductivity in itinerant ferromagnets is formally identical
to the mechanism of appearance of structural order triggered by
the electric polarization in improper ferroelectrics.

Our aim is to establish the uniform phases which are described by the
GL free energy presented in Sec. II. We investigate a quite general GL
model in a situation of a lack of  concrete information about the
values of the parameters of this model for concrete compounds (UGe$_2$,
URhGe, ZrZn$_2$) where the ferromagnetic superconductivity has been
discovered. On the one hand the lack of information makes impossible a
detailed comparison of the theory to  available experimental data but
on the other hand our results are not bound to one or more concrete
substances and can be applied to any unconventional ferromagnetic
superconductor.  In Sec. III the M-trigger effect will be described
when only a linear coupling  of the magnetization $\mbox{\boldmath$M$}$
to the superconducting order parameter $\psi$ is considered in a model
of ferromagnetic superconductors where the spatial dependence of order
parameters and all anisotropy effects are ignored. In Sec. IV we
analyze the influence of quadratic coupling of magnetization to the
superconducting order parameter on the thermodynamics of the
ferromagnetic superconductors. The application of our results to
experimental $(T,P)$ phase diagrams is discussed in Sec. IV.C. In Sec.
V the anisotropy effects are outlined. In Sec. VI we summarize and
discuss our findings.

\section{GINZBURG-LANDAU FREE ENERGY}

The general GL free energy functional, we shall use in our analysis, is
\begin{equation}
\label{eq1} F[\psi,\mbox{\boldmath$M$}]=\int d^3 x f(\psi,
\mbox{\boldmath$M$}),
\end{equation}
where the free energy density $f(\psi,\mbox{\boldmath$M$})$ ( hereafter
called ``free energy'') of a spin-triplet ferromagnetic superconductor
is a sum of five terms~\cite{Machida:2001, Walker:2002, Volovik:1985},
namely,
\begin{equation}
\label{eq2} f(\psi, \mbox{\boldmath$M$}) = f_{\mbox{\scriptsize
S}}(\psi) + f^{\prime}_{\mbox{\scriptsize F}}(\mbox{\boldmath$M$})
+ f_{\mbox{\scriptsize I}}(\psi,\mbox{\boldmath$M$}) +
\frac{\mbox{\boldmath$B$}^2}{8\pi} - \mbox{\boldmath$B.M$}.
\end{equation}
In Eq.~(2) $\psi = \left\{\psi_j;j=1,2,3\right\}$ is a
three-dimensional complex vector describing the superconducting
order and $\mbox{\boldmath$B$} = (\mbox{\boldmath$H$} +
4\pi\mbox{\boldmath$M$}) = \nabla \times \mbox{\boldmath$A$}$ is
the magnetic induction; $\mbox{\boldmath$H$}$ is the external
magnetic field, $\mbox{\boldmath$A$} = \left\{A_j;
j=1,2,3\right\}$ is the magnetic vector potential. The last two
terms on  r.h.s. of Eq.~(2) are related with the magnetic energy
which includes both diamagnetic and paramagnetic effects in the
superconductor; see, e.g., \cite{Vonsovsky:1982, Blount:1979}.

The term $f_{\mbox{\scriptsize S}}(\psi)$ in Eq.~(2) describes the
superconductivity for $\mbox{\boldmath$H$} = \mbox{\boldmath$M$}
\equiv 0$. It can  be written in the form
\begin{equation}
\label{eq3} f_{\mbox{\scriptsize S}}(\psi)= f_{grad}(\psi)
 + a_s|\psi|^2 +\frac{b_s}{2}|\psi|^4 + \frac{u_s}{2}|\psi^2|^2 +
\frac{v_s}{2}\sum_{j=1}^{3}|\psi_j|^4.
\end{equation}
Here
\begin{widetext}
\begin{equation}
\label{eq4} f_{grad}(\psi) = K_1(D_i\psi_j)^{\ast}(D_i\psi_j)+K_2\left[
 (D_i\psi_i)^{\ast}(D_j\psi_j) +  (D_i\psi_j)^{\ast}(D_j\psi_i)\right]
+ K_3(D_i\psi_i)^{\ast}(D_i\psi_i), 
\end{equation}
\end{widetext}
where a summation over the indices $i,j=1,2,3$ is assumed and the
symbol
\begin{equation}
\label{eq5}
 D_j = - i\hbar\frac{\partial}{\partial x_i} + \frac{2|e|}{c}A_j
\end{equation}
of covariant differentiation is introduced. In Eq.~(3), $b_s > 0$
and $a_s = \alpha_s(T-T_s)$, where $\alpha_s$ is a positive
material parameter and $T_s$ is the critical temperature of the
standard second order phase transition which may occur at $H =
{\cal{M}} = 0$; $H =|\mbox{\boldmath$H$}|$, and ${\cal{M}} =
|\mbox{\boldmath$M$}|$. The parameters $u_s$ and $v_s$ describe
the anisotropy of the spin-triplet Cooper pair  and the crystal
anisotropy, respectively, ~\cite{Volovik:1985,Blagoeva:1990}.
  Parameters $K_j$, $(j = 1,2,3)$ in Eq.~(4) are related with the
effective mass tensor of anisotropic Cooper pairs~\cite{Volovik:1985}.

The superconducting part (3) of the free energy $f(\psi, M)$ is derived
from symmetry group arguments and is independent of particular
microscopic models; see, e.g., Refs.~\cite{Volovik:1985, Sigrist:1991}.
According to classifications~\cite{Volovik:1985, Sigrist:1991} the
$p$-wave superconductivity in the cubic point group $O_h$ can be
realized through one-, two-, and three-dimensional representations of
the order parameter. The expressions (3) and (5) incorporate all three
possible cases. The coefficients $b_s$, $u_s$, and $v_s$ in Eq.~(3) are
different for weak and strong spin-orbit couplings but in our
investigation they are considered as undetermined material parameters
which depend on the particular substance.

The  free energy of a standard isotropic ferromagnet is given by
the term $f^{\prime}_{\mbox{\scriptsize F}}(\mbox{\boldmath$M$})$
in Eq.~(2),
\begin{equation}
\label{eq6} f^{\prime}_{\mbox{\scriptsize F}}(\mbox{\boldmath$M$}) =
c_f\sum_{j=1}^{3}|\nabla_j\mbox{\boldmath$M$}_j|^2 +
 a_f(T^{\prime}_f)\mbox{\boldmath$M$}^2 +
 \frac{b_f}{2}\mbox{\boldmath$M$}^4,
\end{equation}
where $\nabla_j = \partial/\partial x_j$ and $b_f > 0$. The
quantity $a_f(T^{\prime}_f) = \alpha_f(T-T^{\prime}_f)$ is
expressed by the material parameter $\alpha_f > 0$ and the
temperature $T^{\prime}_f$ which is different from the critical
temperature $T_f$ of the ferromagnet and this point will be
discussed below. We have already added a negative term ($-2\pi
{\cal{M}}^2$) to the total free energy
$f(\psi,\mbox{\boldmath$M$})$ and that is obvious by setting $H =
0$ in Eq.~(2). The negative energy ($-2\pi{\cal{M}}^2$) should be
added to $f^{\prime}_{\mbox{\scriptsize F}}(\mbox{\boldmath$M$})$.
In this way one obtains the total free energy
$f_{\mbox{\scriptsize F}} (\mbox{\boldmath$M$})$ of the
ferromagnet in a zero external magnetic field that is given by a
modification of Eq.~(6) according to the rule
\begin{equation}
\label{eq7} f_{\mbox{\scriptsize F}} (a_f) =
f^{\prime}_{\mbox{\scriptsize F}} \left[a_f(T^{\prime}_f) \rightarrow
a_f(T_f) \right],
\end{equation}
where  $a_f = \alpha_f (T - T_f)$ and
\begin{equation}
\label{eq8} T_f =  T^{\prime}_f + \frac{2\pi}{\alpha_f}
\end{equation}
is the critical temperature of a standard ferromagnetic phase
transition of second order. This scheme was used in studies of rare
earth ternary compounds~\cite{Vonsovsky:1982, Blount:1979,
Greenside:1981, Ng:1997}. Alternatively~\cite{Kuper:1980}, one may use
from the beginning  the total ferromagnetic free energy
$f_{\mbox{\scriptsize F}}(a_f,\mbox{\boldmath$M$})$ as given by
Eqs.~(6)~-~(8) but in this case the magnetic energy included in the
last two terms on r.h.s. of Eq.~(2) should be replaced with $H^2/8\pi$.
Both approaches are equivalent.

The term
\begin{equation}
\label{eq9} f_{\mbox{\scriptsize I}}(\psi, \mbox{\boldmath$M$}) =
i\gamma_0 \mbox{\boldmath$M$}.(\psi\times \psi^*) + \delta
\mbox{\boldmath$M$}^2 |\psi|^2.
\end{equation}
in Eq.~(2) describes the interaction between the ferromagnetic order
parameter $\mbox{\boldmath$M$}$ and the superconducting order parameter
$\psi$~\cite{Machida:2001,Walker:2002}. The $\gamma_0$-term is the most
substantial for the description of experimentally found ferromagnetic
superconductors~\cite{Walker:2002} while the $\delta
\mbox{\boldmath$M$}^2 |\psi|^2$--term makes the model more realistic in
the strong coupling limit as it gives the opportunity to enlarge the
phase diagram including both positive and negative values of the
parameter $a_s$. In this way  the domain of the stable ferromagnetic
order is extended down to zero temperatures for a wide range of values
of  material parameters and the pressure $P$, a situation that
corresponds to the experiments in ferromagnetic superconductors.

In Eq.~(9) the coupling constant $\gamma_0 >0$ can be represented in
the form $\gamma_0 = 4\pi J$, where $J > 0$ is the ferromagnetic
exchange parameter~\cite{Walker:2002}. In general, the parameter
$\delta$ for ferromagnetic superconductors may take
 both positive and negative values. The values of the material parameters
($T_s$, $T_f$, $\alpha_s$, $\alpha_f$, $b_s$, $u_s$, $v_s$, $b_f$,
$K_j$, $\gamma_0$ and $\delta$) depend on the choice of the concrete
substance and on thermodynamic parameters as temperature $T$ and
pressure $P$.

It is not easy to investigate straightforwardly the  total free
energy (2). In Ref.~\cite{Walker:2002} the authors used  the
criterion~\cite{Abrikosov:1957} for the stability of vortex state
near the phase transition line $T_{c2}(H)$ (see also,
Ref.~\cite{Lifshitz:1980}) and applied it with respect to the
magnetization ${\cal{M}}$ when $H = 0$ for small values of
$|\psi|$ near the phase transition line $T_{c2}({\cal{M}})$. We
are interested in the uniform phases when the order parameters
$\psi$ and $\mbox{\boldmath$M$}$ do not depend on the spatial
vector $\mbox{\boldmath$x$}\in V$ ($V$ is the volume of the
superconductor).
  Therefore, we present a detailed investigation  of the coexistence of
 Meissner superconductivity and ferromagnetic order and,
in particular, we show that the main properties of the uniform phases
can be described when the crystal anisotropy is ignored. We claim that
some of the main features of the uniform phases in unconventional
ferromagnetic superconductors can be reliably outlined even when the
Cooper pair anisotropy is neglected.

The magnetization $\mbox{\boldmath$M$}$ can be always assumed uniform
outside a quite close vicinity of the magnetic phase transition when
 the superconducting order parameter $\psi$ is also uniform, i.e.,
vortex phases are not present in the respective temperature domain.
These conditions are directly satisfied in type I superconductors but
in type II superconductors the temperature should be sufficiently low
and the external magnetic field should be zero.  Nevertheless, in type
II superconductors these requirements  for the appearance of uniform
superconducting states may turn insufficient  in materials having very
high values of the spontaneous magnetization. In this case the uniform
(Meissner) superconductivity and, hence, its coexistence with uniform
ferromagnetic order may not occur even at zero temperature. Up to now
type I unconventional ferromagnetic superconductors are not found
experimentally. The predominant amount of experimental data for
UGe$_2$, URhGe, and ZrZn$_2$ do not give the possibility to conclude
definitely either about the absence or the presence of uniform
superconducting states at low and ultra-low temperatures but recently,
an experimental evidence of uniform coexistence of superconductivity
and ferromagnetism in UGe$_2$ has been reported~\cite{Kotegawa:2004}.

 If real materials can be modelled by the general GL free
energy (1)~-~(9), their ground state properties will be described by
uniform states. The problem about the availability of such states in
real materials at finite temperatures is quite subtle at the present
stage of experimental research. We shall assume that uniform phases can
exist in some unconventional ferromagnetic superconductors, moreover
these phases are solutions of the GL equations corresponding to the
free energy (1)~-~(9). These arguments completely justify our study.

In case of a strong easy axis type of magnetic anisotropy, as is in
UGe$_2$~\cite{Saxena:2000}, the overall complexity of mean-field
analysis of the free energy $f(\psi, \mbox{\boldmath$M$})$ can be
avoided by doing an Ising-like description: $\mbox{\boldmath$M$} =
(0,0,{\cal{M}})$, where ${\cal{M}} = \pm |\mbox{\boldmath$M$}|$ is the
magnetization along the $z$-axis. Because of the thermodynamic
equivalence of  up and down physical states $(\pm \mbox{\boldmath$M$})$
the analysis can be done only for ${\cal{M}} \geq 0$. But this approach
can be  also supported without attracting crystal anisotropy arguments.
When the symmetry of magnetic order is continuous, the symmetry of the
total free energy $f(\psi, \mbox{\boldmath$M$})$ with respect to
$\mbox{\boldmath$M$}$  comes into play and we can avoid the
consideration of equivalent thermodynamic states that occur as a result
of the respective symmetry breaking at the phase transition point but
have no effect on thermodynamics of the system. In the isotropic system
one may again choose the magnetization vector to point in the same
direction as  $z$-axis ($|\mbox{\boldmath$M$}| = M_z = {\cal{M}}$) and
this will not influence the generality of thermodynamic analysis. Here
we prefer an alternative description for which
 the ferromagnetic state can occur as two thermodynamically
equivalent up and down domains with magnetizations $ {\cal{M}}$ and ($
-{\cal{M}}$), respectively.

We shall make the mean-field analysis of the uniform phases and the
possible phase transitions between such phases in a zero external
magnetic field ($\mbox{\boldmath$H$}=0)$ when the crystal anisotropy is
neglected ($v_s \equiv 0$).  The calculations will be  more easy to
understand if we use notations that reduce the number of
 parameters in $f(\psi, \mbox{\boldmath$M$})$ by  introducing
\begin{equation}
\label{eq10} b = (b_s + u_s + v_s).
\end{equation}
Then we redefine the order parameters and all other parameters in the
following way:
\begin{eqnarray}
\label{eq11} &&\varphi_j =b^{1/4}\psi_j = \phi_je^{i\theta_j}\:,\;\;\;
M = b_f^{1/4}{\cal{M}}\:,\\ \nonumber && r =
\frac{a_s}{\sqrt{b}}\:,\;\;\; t =\frac{a_f}{\sqrt{b_f}}\:,
\;\;\; w = \frac{u_s}{b}\:, \;\;\; v =\frac{v_s}{b}\:, \\
\nonumber &&\gamma= \frac{\gamma_0}{b^{1/2}b_f^{1/4}}\:,\;\;\;
\gamma_1= \frac{\delta}{(bb_f)^{1/2}}.
\end{eqnarray}

With the help of Eqs.~(10)~-~(11) and using the uniformity of $\psi$
and $\mbox{\boldmath$M$}$ we write the free energy density $f(\psi,M) =
F(\psi,M)/V$,  in the form
\begin{widetext}
\begin{eqnarray}
\label{eq12} f(\psi,M)& = & r\phi^2 + \frac{1}{2}\phi^4
  + 2\gamma\phi_1\phi_2 M \mbox{sin}(\theta_2-\theta_1) + \gamma_1 \phi^2 M^2
+ tM^2 + \frac{1}{2}M^4 \\ \nonumber && -2w
\left[\phi_1^2\phi_2^2\mbox{sin}^2(\theta_2-\theta_1)
 +\phi_1^2\phi_3^2\mbox{sin}^2(\theta_1-\theta_3) +
 \phi_2^2\phi_3^2\mbox{sin}^2(\theta_2-\theta_3)\right]
 -v[\phi_1^2\phi_2^2 + \phi_1^2\phi_3^2 + \phi_2^2\phi_3^2].
\end{eqnarray}
\end{widetext}
In the above expression the order parameters $\psi$ and
$\mbox{\boldmath$M$}$ are defined per unit volume.

The equilibrium phases are obtained from the equations of state
\begin{equation}
\label{eq13} \frac{\partial f(\mu_0)}{\partial \mu_{\alpha}} = 0,
 \end{equation}
 where
 $\mu = \left\{\mu_\alpha\right\}=
 (M, \phi_1,..., \phi_3,$ $ \theta_1,..., \theta_3)$ and $\mu_0$ denotes an
equilibrium phase. The stability matrix $\tilde{F}$ of the phases
$\mu_0$
 is given by
\begin{equation}
\label{eq14}
 \hat{F}(\mu_0)= \left\{F_{\alpha\beta}(\mu_0)\right\} = \frac{\partial^2f(\mu_0)}
{\partial\mu_{\alpha}\partial\mu_{\beta}}.
\end{equation}

An alternative treatment can be done in terms of real
($\psi^{\prime}_j$) and imaginary ($\psi^{\prime\prime}_j$) parts
of the complex numbers $\psi_j = \psi_j^{\prime} +
i\psi_j^{\prime\prime}$. The calculation with  moduli $\phi_j$ and
phase angles $\theta_j$ of $\psi_j$  is more simple but
 in cases of strongly degenerate phases  some of
the angles $\theta_j$ remain unspecified. Then  an alternative analysis
with the help of the components $\psi_j^{\prime}$ and
$\psi_j^{\prime\prime}$ should be done.

The thermodynamic stability of the phases that are solutions of
Eqs.~(13) is checked with the help of the matrix~(14). An additional
stability analysis is done by the comparison of  free energies of
phases that  satisfy (13) and render the stability matrix (14) positive
in one and the same domain of parameters $\{r,t,\gamma,\gamma_1,w,v\}$.
This step is important because the complicated form of the free energy
generates a great number of solutions of Eqs.~(13) and we have to sift
out the stable from metastable phases that correspond either to global
or local
 minima of the free energy, respectively ~\cite{Uzunov:1993}.

 Some solutions  of Eqs.~(13) have  a marginal stability, i.e., their
 stability matrix (14) is
  neither positively nor negatively definite.
 This is often a result of the degeneration of
phases with broken continuous symmetry. If the reason for the lack
of a clear positive definiteness of the stability matrix is
precisely the mentioned degeneration of the ground state, one may
reliably conclude that the respective phase is stable. If there is
another reason, the analysis of the matrix (14) will be
insufficient
 to determine the respective stability property. These
cases are quite rare and occur for particular values of the parameters
$\{r,t,\gamma,...\}$.

\section{SIMPLE CASE OF M-TRIGGERED SUPERCONDUCTIVITY}

We shall consider the Walker-Samokhin model~\cite{Walker:2002} when
only the $M\phi_1\phi_2-$coupling between the order parameters $\psi$
and $M$ is taken into account ($\gamma > 0$, $\gamma_1 = 0$) and  the
anisotropies $(w=v=0)$ are ignored. The uniform phases and the phase
diagram in this case were investigated in
Refs.~\cite{Shopova1:2003,Shopova2:2003, Shopova3:2003}. Here we
summarize the main results in order to make a clear comparison with the
new results presented in Sections~IV and V. Our main aim is the
description of {\em a trigger effect} which consists of the appearance
of a ``compelled superconductivity'' caused by the presence of
ferromagnetic order (here, this is a standard uniform ferromagnetic
order); see also Refs.~\cite{Shopova1:2003,Shopova2:2003,
Shopova3:2003} where this effect has been already established and
briefly discussed. As mentioned in the Introduction, a similar trigger
effect is known in the physics of improper ferroelectrics. We shall set
$\theta_3 \equiv 0$ and use the notation $\theta \equiv \Delta\theta =
(\theta_2 - \theta_1)$.

\subsection{Phases}

The possible (stable, metastable and unstable)
 phases are given in Table 1 together with the respective
existence and stability conditions.  The normal or disordered
phase, denoted in Table 1 by $N$, always exists (for all
temperatures $T \geq 0)$ and is stable for $t >0$, $r > 0$. The
superconducting phase denoted in Table 1 by SC1 is unstable. The
same is valid for the phase of coexistence of ferromagnetism and
superconductivity denoted in Table 1 by CO2. The N--phase, the
ferromagnetic phase (FM), the superconducting phases (SC1--3) and
two of the phases of coexistence (CO1--3) are generic phases
because they appear also in the decoupled case $(\gamma\equiv 0)$.
When the $M\phi_1\phi_2$--coupling is not present, the phases
SC1--3 are identical and represented by the order parameter $\phi$
with components $\phi_j$ that participate on  equal footing. The
asterisk attached to the stability condition of the second
superconductivity phase (SC2) indicates that our analysis is
insufficient to determine whether this phase corresponds to a
minimum of the free energy. It will be shown that the phase SC2,
two other purely superconducting phases and the coexistence phase
CO1, have no chance to become stable for $\gamma \neq 0$. This is
so, because the phase of coexistence of superconductivity and
ferromagnetism (FS in Table 1), that does not occur for $\gamma =
0$, is stable and has a lower free energy in their domain of
stability.  A second domain $(M < 0)$ of the FS phase exists and
is denoted in Table 1 by FS$^*$. Here we shall describe only the
first domain FS. The domain FS$^{\ast}$ is considered in the same
way.

\begin{table*}
\caption{\label{tab:table1}Phases and their existence and stability
properties [$\theta = (\theta_2-\theta_1)$, $k = 0, \pm 1,...$].}
\begin{ruledtabular}
\begin{tabular}{llll}
 Phase & order parameter & existence conditions & stability domain \\ \hline\hline

N & $\phi_j = M = 0$ & always & $t > 0, r > 0$ \\ \hline

FM & $\phi_j = 0$, $M^2 = -t$& $t < 0$& $r>0$, $r > r_e(t)$\\ \hline

SC1 & $\phi_1=M=0$, $\phi^2 = -r$ & $r<0$ & unstable  \\ \hline

SC2 & $\phi^2= -r$, $\theta = \pi k$, $M = 0$ & $r<0$ & $(t > 0)^*$\\
\hline

SC3 & $\phi_1=\phi_2=M=0$, $\phi^2_3 = -r$ & $r<0$ &$r<0$, $t>0$\\
\hline

CO1 &$\phi_1= \phi_2=0$, $\phi^2_3 = -r$, $M^2=-t$&$r<0$, $t<0$ &
$r<0$, $t<0$ \\  \hline

CO2 &$\phi_1=0$, $\phi^2 = -r$, $\theta=\theta_2=\pi k$, $M^2=-t$&
$r<0$, $t<0$ &  unstable \\ \hline

FS & $2\phi_1^2 = 2\phi_2^2 = \phi^2 = -r + \gamma M$, $\phi_3 = 0$ &
$\gamma M > r$ &  $3M^2>(-t +\gamma^2/2)$ \\
& $\theta= 2\pi(k - 1/4) $, $\gamma r = (\gamma^2-2t)M - 2M^3$ & & $M >
0$ \\ \hline

FS$^{\ast}$ & $2\phi_1^2 = 2\phi_2^2 = \phi^2 = -(r + \gamma M)$,
$\phi_3 = 0$ & $-\gamma M > r$ &  $3M^2>(-t +\gamma^2/2)$ \\

& $\theta= 2\pi(k + 1/4) $,
$\gamma r = (2t -\gamma^2)M + 2M^3$ & & $M < 0$ \\
\end{tabular}
\end{ruledtabular}
\end{table*}

\begin{figure*}
\includegraphics{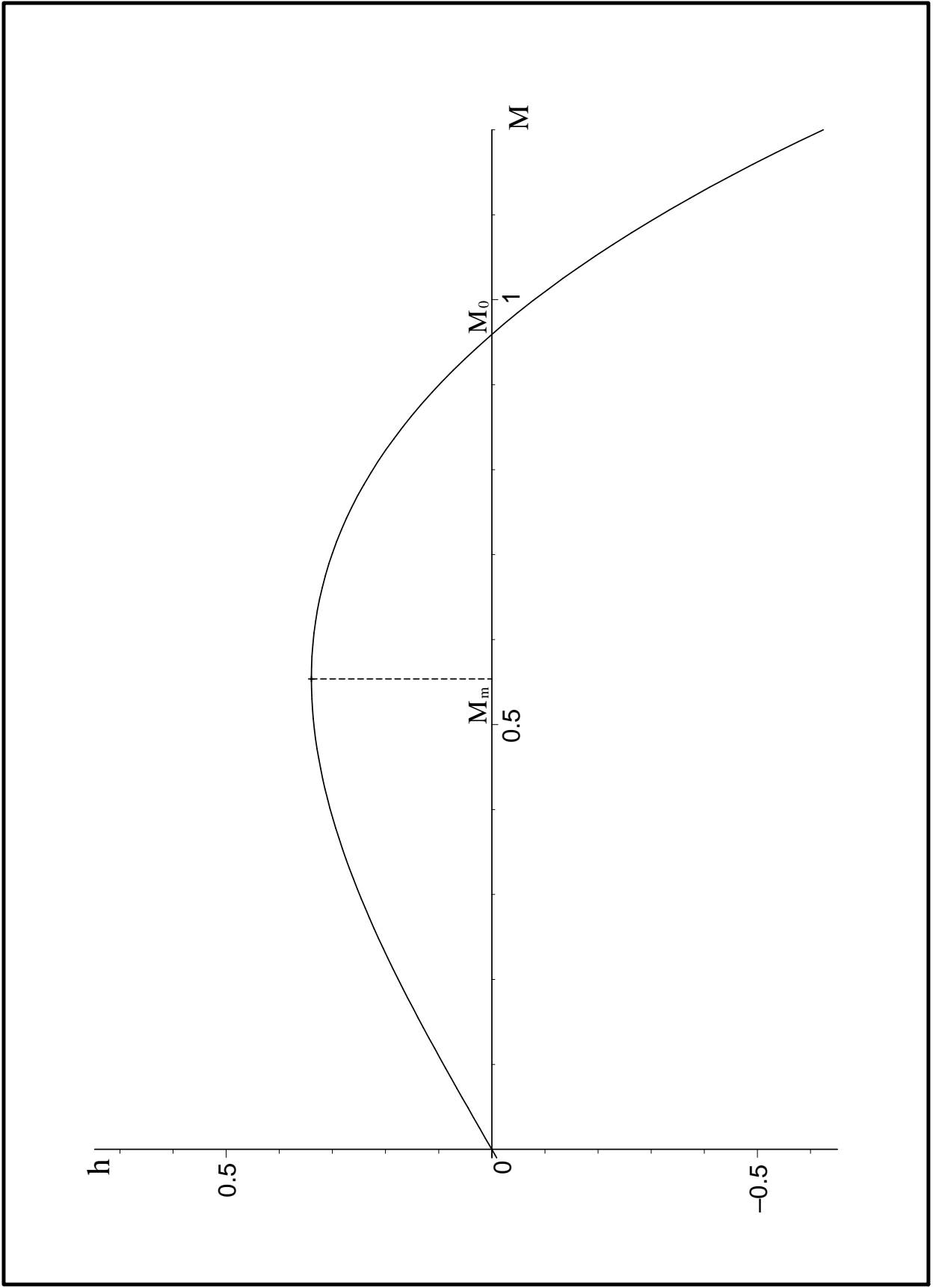}
\caption{\label{fig:wide}$h=\gamma r/2$ as a function of $M$ for
$\gamma = 1.2$, and $t = -0.2$. The parameters $r$, $t$, and $\gamma$
are given by Eq.~(11).}
\end{figure*}

\begin{figure*}
\includegraphics{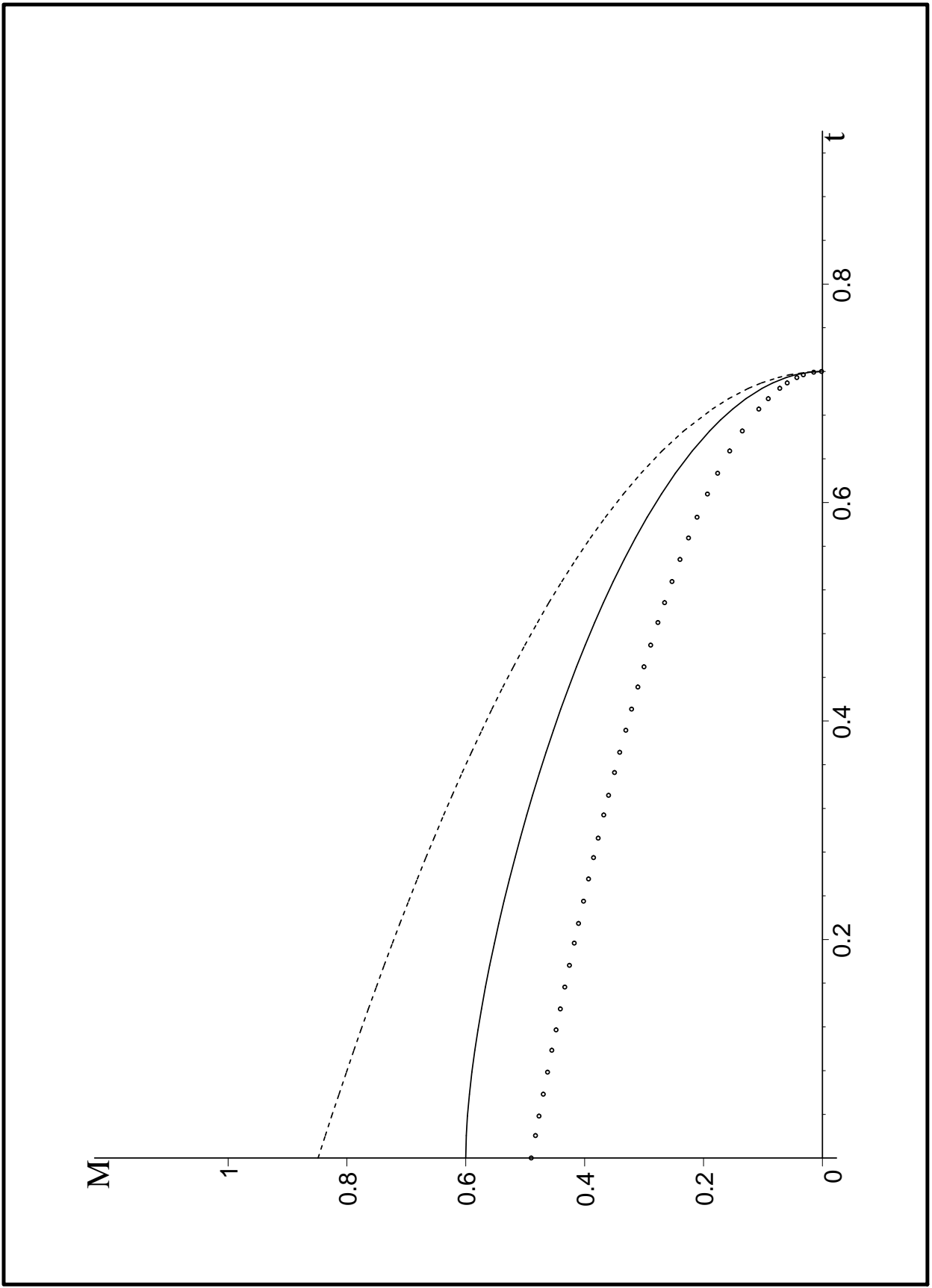}
\caption{\label{fig:wide}The magnetization $M$ versus $t$ for $\gamma =
1.2$: the dashed line represents $M_0$, the solid line represents
$M_{eq}$, and the dotted line corresponds to $M_m$.}
\end{figure*}

\begin{figure*}
\includegraphics{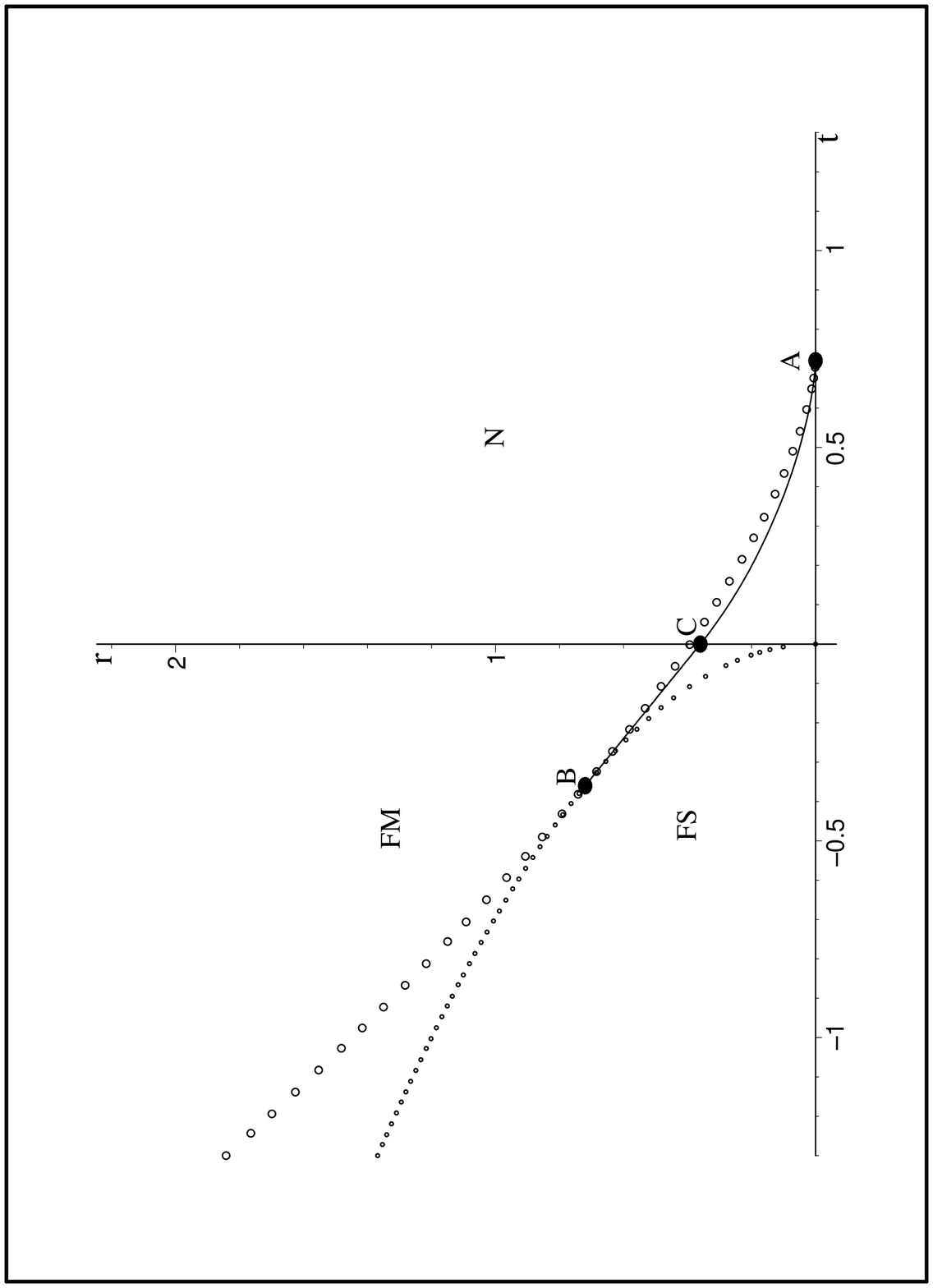}
\caption{\label{fig:wide}The phase diagram in the plane ($t$, $r$) with
two tricritical points (A and B) and a triple point $C$; $\gamma =
1.2$. The parameters $r \sim [T - T_s(P)]$ and $t \sim [T-T_f(P)]$ are
defined by Eq.~(11). The domains of existence and stability of the
phases N, FM and FS are shown. The line of circles represents the
function $r_m(t)$ given by Eq.~(17). The dotted line represents the
function $r_e(t)$ given by Eq.~(15). On the left of point $B$, the same
dotted curve corresponds to a FM-FS phase transition of second order.
The equilibrium lines of N-FS and FM-FS phase transitions of first
order are given by the solid lines $AC$ and $CB$, respectively.}
\end{figure*}

The cubic equation for magnetization of  FS-phase (see Table~1) is
shown in Fig.~1 for $\gamma = 1.2$ and $t = -0.2$. For any $\gamma
> 0$ and $t$, the stable FS thermodynamic states are given by $r
(M) < r_m = r(M_m)$ for $M > M_m > 0$, where $M_m$ corresponds to
the maximum of the function $r(M)$. The dependance of $M_m(t)$ and
$M_0(t) = (-t + \gamma^2/2)^{1/2} = \sqrt{3}M_m(t)$ on $t$ is
drawn in Fig.~2 for $\gamma = 1.2$.  Functions $r_m(t) =
4M_m^3(t)/\gamma$ for $t < \gamma^2/2$ (depicted by the line of
circles in Fig.~3) and
\begin{equation}\label{eq15}
  r_e(t) = \gamma|t|^{1/2},
\end{equation}
 for $t < 0$  define the borderlines of
stability and existence of FS.

\subsection{Phase diagram}

We have outlined the domain in the ($t$, $r$) plane where the FS phase
exists and is a minimum of the free energy. For $r < 0$ the cubic
equation  for $M$ (see Table 1) and the existence and stability
conditions are satisfied for any $M \geq 0$ provided $t \geq \gamma^2
$. For $ t < \gamma^2$ the condition $M \geq M_0$ have to be fulfilled,
here the value
 $M_0 = (-t + \gamma^2/2)^{1/2}$ of $M$ is obtained from $r(M_0) = 0$. Thus
for $r = 0$ the N-phase is stable for
 $t \geq \gamma^2/2$, and FS is stable for $t \leq \gamma^2/2$.
For $r > 0$, the requirement for the stability of FS leads to the
inequalities
\begin{equation}
\label{eq16}
  max\left(\frac{r}{\gamma}, M_m\right) < M < M_0,
\end{equation}
where $M_m = (M_0/\sqrt{3})$ and $M_0$ should be the positive solution
of the cubic equation of state from Table~1; $M_m > 0$ gives a maximum
of the function $r(M)$; see also Figs.~1 and 2.

The further analysis  defines the existence and  stability domain of FS
below the line AB denoted by circles (see Fig.~3). In Fig.~3 the curve
of circles starts from the point A with coordinates ($\gamma^2/2$, $0$)
and touches two other (solid and dotted) curves at the point B with
coordinates ($t_B=-\gamma^2/4$, $r_B= \gamma^2/2$).  Line of
 circles represents the function
$r(M_m) \equiv r_m(t)$ where
\begin{equation}
\label{eq17}
 r_m(t) = \frac{4}{3\sqrt{3}\gamma} \left (\frac{\gamma^2}{2} -
 t\right)^{3/2}.
\end{equation}
Dotted line represents  $r_e(t)$, defined by Eq.~(15). The inequality
$r < r_m(t)$ is a condition for the stability of FS, whereas the
inequality $r \leq r_e(t)$ for $ (-t) \geq \gamma^2/4$ is a condition
for the existence of FS as a solution of the respective equation of
state. This existence condition for FS is obtained from $\gamma M
> r$ (see Table 1).

In the region on the left of the point B in Fig.~3, the FS phase
satisfies the existence condition $\gamma M > r$ only
 below the dotted line. In the domain confined between the lines of circles
 and the dotted
line on the left of the point B the stability condition for FS is
satisfied but the existence condition is broken. The inequality $r \geq
r_e(t)$ is the stability condition of FM for $ 0 \leq (-t) \leq
\gamma^2/4$. For $(-t) > \gamma^2/4$ the FM phase is stable for all $r
\geq r_e(t)$.

In the region confined by the line of circles AB, the dotted line for $
0 < (-t) < \gamma^2/4$, and the $t-$axis, the phases N, FS and FM have
an overlap of stability domains. The same is valid for FS, the SC
phases and CO1 in the third quadrant of the plane ($t$, $r$). The
comparison of the respective free energies for $r < 0$ shows that the
stable phase is FS whereas the other phases are metastable within their
domains of stability.

The part of
 the $t$-axis given by $r=0$ and $t > \gamma^2/2$
 is a phase transition line of second order
which describes the N-FS transition. The same transition
 for $0 < t < \gamma^2/2$ is represented by the solid line AC which
is the equilibrium transition line of a first order phase transition.
The equilibrium transition curve is given by the function
\begin{equation}
\label{eq18}
 r_{eq}(t) =
\frac{1}{4}\left[3\gamma - \left(\gamma^2 + 16t
\right)^{1/2}\right]M_{eq}(t).
\end{equation}
Here
  \begin{equation}
\label{eq19}
 M_{eq}(t) =
\frac{1}{2\sqrt{2}}\left[\gamma^2 - 8t + \gamma\left(\gamma^2 +
 16t \right)^{1/2}\right]^{1/2}
\end{equation}
is the equilibrium jump of the magnetization. The order of the N-FS
transition changes at the tricritical point A.

The domain above the solid line AC and below the line of circles for $
t > 0$ is the region of a possible
 overheating of FS.
The domain of overcooling of the N-phase is confined by the solid line
AC and the axes ($t > 0$, $r >0$). At the triple point C with
coordinates
 [0, $r_{eq}(0) = \gamma^2/4$]
the phases N, FM, and FS coexist. For $t < 0$ the straight line
\begin{equation}
\label{eq20} r_{eq}^* (t) =  \frac{\gamma^2}{4} + |t|,\;\;\;\;\;\; t_B
< t < 0,
\end{equation}
describes the extension of the equilibrium phase transition line of the
N-FS first order transition to negative values of $t$.
 For $t < t_B$
 the equilibrium phase transition FM-FS is of second order and is
given by the dotted line on the left of the point B which is the second
tricritical point in this phase diagram. Along the first order
transition line
 $r_{eq}^{\ast}(t)$ given by~ Eq.~(\ref{eq20}) the equilibrium value
 of $M$ is $M_{eq} =\gamma/2$,  which
implies an equilibrium order parameter jump at the FM-FS transition
equal to ($\gamma/2 - \sqrt{|t|}$). On the dotted line of the second
order FM-FS
 transition the equilibrium value
of $M$ is equal to that of the FM phase ($M_{eq} = \sqrt{|t|}$).
The FM phase does not exist below $T_s$ and this is a shortcoming
of the model~(\ref{eq12}) with $\gamma_1 = 0$.

The equilibrium  FM-FS and N-FS phase transition lines in Fig.~3 can be
expressed by the respective equilibrium phase transition temperatures
$T_{eq}$ defined by the equations $r_e = r(T_{eq})$, $r_{eq} =
r(T_{eq})$, $r^{\ast}_{eq} = r(T_{eq})$, and with the help of the
relation $M_{eq} = M(T_{eq})$. This limits  the possible variations of
parameters of the theory.
 For example, the critical temperature
($T_{eq} \equiv T_c$) of the FM-FS second order transition
 ($\gamma^2/4 < -t$)  is obtained in the form
$T_{c} = (T_s + 4\pi J{\cal{M}}/\alpha_s)$, or, using ${\cal{M}} =
(-a_f/b_f)^{1/2}$,
\begin{equation}
\label{eq21} T_{c} = T_s -\frac{T^{\ast}}{2} + \left[
\left(\frac{T^{\ast}}{2}\right)^2 + T^{\ast}(T_f-T_s)\right]^{1/2}.
\end{equation}
Here $T_f > T_s$, and $T^{\ast} = (4\pi J)^2\alpha_f/\alpha_s^2b_f$ is
 a characteristic temperature of the model~(\ref{eq12}) with
 $\gamma_1=w=v=0$. A discussion of Eq.~(21) is given in Sec. IV.C.

The investigation of the conditions for the validity of
Eq.~(\ref{eq21}) leads to the conclusion that the FM-FS continuous
phase transition (at $\gamma^2 < -t)$ will be possible only if the
following condition is satisfied:
\begin{equation}
\label{eq22} T_{f} - T_s > \ = (\varsigma +
\sqrt{\varsigma})T^{\ast},
\end{equation}
where $\varsigma = b_f\alpha_s^2/4b_s\alpha_f^2$.
 Therefore, the second
order FM-FS transition should disappear for a sufficiently large
$\gamma$--coupling. Such a condition does not exist for the first order
transitions FM-FS and N-FS.

 The inclusion of the gradient term (4) in the free
energy~(\ref{eq2}) should lead to a depression of the equilibrium
transition temperature. As the magnetization increases with the
decrease of the temperature, the vortex state should occur at
temperatures which are lower than
 the equilibrium temperature $T_{eq}$ of
the Meissner state. For example, the critical temperature
($\tilde{T}_c$)
 corresponding to the vortex phase
of FS-type has been evaluated~\cite{Walker:2002} to be
 lower than the critical temperature ~(\ref{eq21}): $(T_c - \tilde{T}_c) =
4\pi \mu_B{\cal{M}}/\alpha_s$, where $\mu_B = |e|\hbar/2mc$ is the Bohr
magneton.
 For $J \gg \mu_B$, we have $T_c \approx \tilde{T}_c$.

For $ r > 0$, namely, for temperatures $T > T_s$ the superconductivity
is triggered by the magnetic order through the $\gamma$-coupling. The
superconducting phase for $T > T_s$ is entirely in the $(t,r)$ domain
of the ferromagnetic phase. Therefore, the uniform supeconducting phase
can occur for $T > T_s$ only through a coexistence with the
ferromagnetic order.

In the next Sections we shall focus on the temperature range $T > T_s$
which seems to be of main practical interest. We shall not dwell on the
superconductivity in the fourth quadrant
 $(t >0,r<0)$ of the $(t,r)$ diagram where pure superconducting phases
can occur for systems with $T_s > T_f$,  but this is not the case
for UGe$_2$, URhGe and ZrZn$_2$. Also we shall not discuss the
possible metastable phases in the third quadrant $(t<0,r<0)$ of
the $(t,r)$ diagram.

\subsection{Magnetic susceptibility}

We consider the longitudinal magnetic susceptibility $\chi_1 =
(\chi_{\mbox{\scriptsize V}}/V)$ per unit volume~\cite{Shopova3:2003}.
The external magnetic field $\mbox{\boldmath$H$} = (0,0,H)$ with $ H =
\left(\partial f/\partial {\cal{M}}\right)$ has the same direction as
the magnetization $\mbox{\boldmath$M$}$. We shall calculate the
quantity $\chi = \sqrt{b_f}\chi_1$ for the equilibrium thermodynamic
states $\mu_0$ given by Eq.~(13). Having in mind the relations (11)
between $M$ and ${\cal{M}}$, and between $\psi$ and $\varphi$ we can
write
\begin{equation}
\label{eq23} \chi^{-1} =  \frac{d}{d
M_0}\left[\left(\frac{\partial f} {\partial
M}\right)_{T,\varphi_j} \right]_{\mu_0},
\end{equation}
where the equilibrium magnetization $M_0$ and equilibrium
superconducting order parameter components $\varphi_{0j}$ should be
taken for the respective equilibrium phase. See Table~1, where the
suffix ``0'' of $\phi$, $\theta$, and $M$ is omitted; hereafter this
suffix will be often omitted. The value of the equilibrium
magnetization $M$ in FS is the maximal nonnegative root of the cubic
equation in $M$ given in Table~1.

 From Eq.~(23) we obtain the susceptibility $\chi$ of  FS phase in
the form
\begin{equation}
\label{eq24} \chi^{-1} = -\gamma^2 +  2t + 6M^2.
\end{equation}
The susceptibility of the other phases has the usual expression
\begin{equation}
\label{eq25} \chi^{-1} =  2t + 6M^2.
\end{equation}
Eq.~(25) yields as results  the paramagnetic susceptibility
  ($\chi_P = 1/2t$; $t>0$) of the normal phase and the
ferromagnetic susceptibility ($\chi_F = 1/4|t|$; $t <0$) of FM. These
susceptibilities can be compared with the susceptibility $\chi$ of FS
which cannot be calculated analytically in the whole domain of
stability of FS.  Therefore, we shall consider the close vicinity of
the N-FS and FM-FS phase transition lines.

Near the second order phase transition line on the left of the point
$B$ ($t < t_B$), the magnetization has a smooth behavior and the
magnetic susceptibility does not exhibit any singularities like jump or
divergence. For $t > \gamma^2/2$, the magnetization is given by $M =
(s_- + s_+)$, where
\begin{equation}
\label{eq26}
 s_{\pm} =\left\{- \frac{\gamma r}{4} \pm \left[
\frac{(t-\gamma^2/2)^3}{27} + \left( \frac{\gamma
r}{4}\right)^2\right]^{1/2} \right\}^{1/3}.
\end{equation}
When $r = 0$, it is obvious that also $M = 0$.  For $|\gamma r| \ll (t
- \gamma^2/2)$  we have $M \approx -\gamma r/ (2t-\gamma^2) \ll 2t$.
Therefore, in a close vicinity $(r < 0)$ of $r = 0$ along the second
order phase transition line $(r=0, t>\gamma^2/2)$ the magnetic
susceptibility is well described by the paramagnetic law $\chi_P =
(1/2t)$. For $r< 0$ and $t \rightarrow \gamma^2/2$, we obtain $M =
-(\gamma r/2)^{1/3}$ which gives
\begin{equation}
\label{eq27}
 \chi^{-1} =  6\left(\frac{\gamma |r|}{2}\right)^{2/3}.
\end{equation}

On the phase transition line $AC$
\begin{equation}
\label{eq28} M_{eq}(t) = \frac{1}{2\sqrt{2}}\left[\gamma^2 - 8t +
\gamma\left(\gamma^2 + 16t \right)^{1/2}\right]^{1/2}
\end{equation}
and, hence,
\begin{equation}
\label{eq29}
 \chi^{-1} = -4t - \frac{\gamma^2}{4}\left[1 -3 \left( 1 +
 \frac{16t}{\gamma^2}\right)^{1/2}\right].
\end{equation}
At the tricritical point $A$ this result gives $\chi^{-1}(A) = 0$, and
at the triple point $C$ with coordinates ($0$, $\gamma^2/4$) we have
$\chi(C) = (2/\gamma^2)$. On the line $BC$ we obtain $M=\gamma/2$ so
\begin{equation}
\label{eq30} \chi^{-1} = 2t + \frac{\gamma^2}{2}.
\end{equation}
At the tricritical point $B$ with coordinates ($-\gamma^2/4$,
$\gamma^2/2$) the result is $\chi^{-1}(B)= 0$.

To investigate the magnetic susceptibility tensor we shall consider
arbitrary orientations of the vectors $\mbox{\boldmath$H$}$ and
$\mbox{\boldmath$M$}$. We  denote the spatial directions
$(\mbox{\boldmath$x$},\mbox{\boldmath$y$},\mbox{\boldmath$z$})$ by
$(1,2,3)$.

The components of the inverse magnetic susceptibility tensor
\begin{equation}
\label{eq31} \hat{\chi}^{-1}_1 =\hat{\chi}^{-1}\sqrt{b_f} =
\left\{\chi^{-1}_{ij}\right\} \sqrt{b_f}
\end{equation}
can be represented in the form
\begin{equation}
\label{eq32} \chi^{-1}_{ij} = 2(t + M^2)\delta_{ij} + 4M_iM_j +
i\gamma\frac{\partial}{\partial
M_j}(\varphi\times\varphi^{\ast})_i,
\end{equation}
where $M$ and $\varphi_j$ are taken at their equilibrium values: $M_0$,
$\varphi_{0j}$, $\theta_{0j}$. The last term in r.h.s. of Eq.~(29) is
equal to zero for all phases in Table~1 except for FS and FS$^{\ast}$.
When the second term in Eq.~(30) is equal to zero we obtain the known
result of the susceptibility tensor for second order phase transitions;
see, e.g.,~\cite{Uzunov:1993}.

In FS phase $\phi_{j}$ depend on $M_j$ and we can choose again
$\mbox{\boldmath$M$} = (0,0,M)$ and use the results from Table~1 for
the equilibrium values of $\phi_j$, $\theta$ and $M$. Then the
components $\chi^{-1}_{ij}$ corresponding to FS are
\begin{equation}
\label{eq33} \chi^{-1}_{ij} = 2(t + M^2)\delta_{ij} + 4M_iM_j
-\gamma^2\delta_{i3}.
\end{equation}
Thus we have $\chi^{-1}_{i\neq j}= 0$,
\begin{equation}
\label{eq34} \chi^{-1}_{11} = \chi^{-1}_{22} = 2(t + M^2),
\end{equation}
and $\chi^{-1}_{33}$  coincides with the inverse longitudinal
susceptibility
 $\chi^{-1}$ as given by  Eq.~(24).

 \subsection{Entropy and specific heat}

The entropy $S(T) \equiv (\tilde{S}/V) = -V\partial (f/\partial T) $
and the specific heat $C(T) \equiv (\tilde{C}/V) = T(\partial
S/\partial T)$ per unit volume $V$ are calculated in a standard
way~\cite{Uzunov:1993}. We are interested in the jumps of these
quantities on the N-FM, FM-FS, and N-FS transition lines. The behavior
of $S(T)$ and $C(T)$ near the N-FM phase transition and near the FM-FS
phase transition line of second order on the left of the point $B$
(Fig.~3) is known from the standard theory of critical phenomena  and
for this reason we focus our attention on the first order phase
transitions  FS-FM and FS-N for $t>-\gamma^2/4$, i.e., on the right of
the point $B$ in Fig.~3.

We make use of the equations for the order parameters $\psi$ and $M$
from Table~1 and apply the standard procedure for the calculation of
$S$:
\begin{equation}
\label{eq35}
 S(T) = - \frac{\alpha_s}{\sqrt{b_s}}\phi^2 -
\frac{\alpha_f}{\sqrt{b_f}}M^2.
\end{equation}
The next step is to calculate the entropies $S_{ FS}(T)$ and $S_{FM}$
of the ordered phases FS and FM. We shall stick to the usual convention
$F_{N} = Vf_{N}=0$ for the free energy of the N-phase , so we must set
$S_{N}(T)=0$.

 Near the  second order phase transition line ($r=0$,
$t>\gamma^2/2$), $S_{FS}(T)$ is a smooth function of $T$ and has
no jump but the specific heat $C_{FS}$ has a jump at $T=T_s$, i.e.
for $r=0$. This jump is given by
\begin{equation}
\label{eq36}
 \Delta C_{FS}(T_s) = \frac{\alpha_s^2T_s}{b_s}\left[ 1 -
 \frac{1}{1 - 2t(T_s)/\gamma^2}\right].
\end{equation}
The jump $\Delta C_{FS}(T_s)$ is higher than the usual jump $\Delta
C(T_c) = T_c\alpha_s^2/b_s$ known from the Landau theory of standard
second order phase transitions~\cite{Uzunov:1993}.

The entropy jump $\Delta S_{AC}(T) \equiv S_{FS}(T) $ on the line $AC$
is

\begin{eqnarray}
\label{eq37}
 \lefteqn{\Delta S_{AC}(T)
 =}\\ \nonumber &&-M_{eq}\left\{\frac{\alpha_s\gamma}{4\sqrt{b_s}}\left[1 + \left(1 +
 \frac{16t}{\gamma^2}\right)^{1/2}\right]
 -\frac{\alpha_f}{\sqrt{b_f}}M_{eq}\right\},
\end{eqnarray}
where $M_{eq}$ is given by Eq.~(19). From Eqs.~(19) and (37), we have
$\Delta S(t=\gamma^2/2) = 0$, i.e., $\Delta S(T)$ becomes equal to zero
at the tricritical point $A$. We find also from Eqs.~(19) and (37) that
at the triple point $C$ the entropy jump is
\begin{equation}
\label{eq38} \Delta S(t=0) =
  -\frac{\gamma^2}{4}\left(\frac{\alpha_s}{\sqrt{b_s}} +
   \frac{\alpha_f}{\sqrt{b_f}} \right).
\end{equation}

On the line $BC$ the entropy jump is defined by $\Delta S_{BC}(T) =
[S_{FS}(T)-S_{FM}(T)]$. We obtain
\begin{equation}
\label{eq39} \Delta S_{BC}(T) =
 \left( |t| -\frac{\gamma^2}{4}\right)\left(\frac{\alpha_s}{\sqrt{b_s}}
  + \frac{\alpha_f}{\sqrt{b_f}}
   \right).
\end{equation}
At the tricritical point $B$ this jump is equal to zero as  should be.
The calculation of the specific heat jump on the first order phase
transition lines $AC$ and $BC$ is redundant for two reasons. Firstly,
the jump of the specific heat at a first order phase transition differs
from the entropy by a factor of order of unity. Secondly, in caloric
experiments where the relevant quantity is the latent heat $Q = T
\Delta S(T)$, the specific heat jump can hardly be distinguished.

\subsection{Note about a simplified theory}

The analysis in this Section  can be done following an approximate
scheme known from the theory of improper ferroelectrics; see,
e.g., Ref.~\cite{Cowley:1980}.  In this approximation the order
parameter $M$  is considered small enough which makes possible to
ignore $M^4$-term in the free energy. Then one easily obtains from
the data for FS presented in Table~1 or by a direct calculation of
the respective reduced free energy that the order parameters
$\phi$ and $M$ of FS--phase are described by the simple equalities
$r = (\gamma M -\phi^2)$ and $M = (\gamma/2t)\phi^2$.  For
ferroelectrics working with oversimplified free energy
 gives a substantial departure of theory from
experiment~\cite{Cowley:1980}. For ferromagnetic superconductors the
domain of reliability of this approximation could be the close vicinity
of the ferromagnetic phase transition, i.e. for temperatures near  the
critical temperature $T_f$. This discussion can be worthwhile if only
the primary order parameter also exists in the same narrow temperature
domain ($\phi > 0$). Therefore, the application of the simplified
scheme can be useful in systems, where $T_s \ge T_f$.

 For $T_s<T_f$,
the analysis can be simplified if we suppose  a relatively small value
of the modulus $\phi$ of the superconducting order parameter. This
approximation should be valid in some narrow temperature domain near
the line of second order phase transition from FM to FS.

\section{EFFECT OF SYMMETRY CONSERVING COUPLING}

Here we shall include in our consideration   both linear and
quadratic couplings of magnetization to the superconducting order
parameter which means that both parameters $\gamma$ and $\gamma_1$
in free energy~(12) are different from zero. In this way we shall
investigate the effect of the symmetry conserving $\gamma_1$-term
in the free energy on the thermodynamics of the system. When
$\gamma$ is equal to zero but $\gamma_1 \ne 0$ the analysis is
easy and the results are known from the theory of bicritical and
tetracritical points~\cite{Uzunov:1993, Toledano:1987, Liu:1973,
Imry:1975}. For the problem of coexistence of conventional
superconductivity and ferromagnetic order the analysis $(\gamma =
0, \gamma_1 \neq 0)$ was made in Ref.~\cite{Vonsovsky:1982}.

 At this
stage we shall not take into account any anisotropy effects
 because we do not want to obscure the influence of quadratic
interaction by considering too many parameters. For
$\gamma,\gamma_1 \ne
 0$ and $w=0$, $v=0$ the results again  can be presented in an analytical form,
 only a small
part of phase diagram should be calculated numerically.

\subsection{Phases}

The calculations show that for temperatures $T > T_s$, i.e., for $r >
0$, we have again three stable phases. Two of them are quite simple:
the normal ($N$-) phase with existence and stability domains shown in
Table~1, and the FM phase with the existence condition $ t<0$ as shown
in Table~1, and a stability domain defined by the inequality $r_e^{(1)}
\le r$. Here
\begin{equation}
\label{eq40} r_e^{(1)}=\gamma_1t + \gamma\sqrt{-t},
\end{equation}
and one can compare it with the respective expression~(15) for
$\gamma_1=0$. In this paragraph we shall retain the same notations as
in Sec.~III, but with a superscript $(1)$ in order to distinguish them
from the case $\gamma_1=0$ The third stable phase for $r>0$ is a more
complex variant of the mixed phase FS and its domain FS$^*$, discussed
in Sec. III. The symmetry of the FS phase coincides with that found in
~\cite{Walker:2002}.

We have to mention that for $r<0$ there are five pure superconducting
($M =0$, $\phi
> 0$) phases. Two of them, $(\phi_1 > 0, \phi_2 =
\phi_3 =0)$ and $(\phi_1 =0, \phi_2>0, \phi_3>0)$ are unstable. Two
other phases, $(\phi_1>0, \phi_2>0, \phi_3 =0, \theta_2 = \theta_1 +
\pi k)$ and $(\phi_1>0,\phi_2>0, \phi_3>0, \theta_2 = \theta_1 + \pi k,
\theta_3$ -- arbitrary; $k=0,\pm1,...)$ show a marginal stability for $
t > \gamma_1 r$.

Only one of the five pure superconducting phases,  the phase SC3, given
in Table~1, is stable. In case of $\gamma_1 \neq 0$ the values of
$\phi_j$ and the existence domain of SC3 are the same as shown in Table
~1 for $\gamma_1 =0$ but the stability domain is different and is given
by $t > \gamma_1 r$. When the anisotropy effects are taken into account
the phases exhibiting marginal stability within the present
approximation may become stable. Besides, three other mixed phases $(M
\neq 0, \phi
>0)$ exist for $r < 0$ but one of them is metastable (for
$\gamma_1^2 >1, t < \gamma_1 r$, and $r < \gamma_1 t$) and the
other two are absolutely unstable. Here the thermodynamic behavior
for $r < 0$ is much more abundant in phases than for improper
ferroelectrics with two component primary order parameter
~\cite{Toledano:1987}. However, at this stage of experimental
needs about the properties of unconventional ferromagnetic
superconductors the investigation of the phases for temperatures
$T < T_s$ is not of primary interest and for this reason we shall
focus our attention on the temperature domain $r > 0$.

The FS phase for $\gamma_1 \ne 0$ is described by the following
equations:
\begin{equation}
\label{eq41} \phi_1 = \phi_2=\frac{\phi}{\sqrt{2}}\:, \;\;\;
\phi_3 = 0,
\end{equation}
\begin{equation}
\label{eq42} \phi^2= (\pm \gamma M-r-\gamma_1 M^2),
\end{equation}
\begin{equation}
\label{eq43} (1-\gamma_1^2)M^3\pm \frac{3}{2} \gamma \gamma_1 M^2
+\left(t-\frac{\gamma^2}{2}-\gamma_1 r\right)M \pm \frac{\gamma
r}{2}=0,
\end{equation}
and
\begin{equation}
\label{eq44} (\theta_2 - \theta_1) = \mp \frac{\pi}{2} + 2\pi k,
\end{equation}
($k = 0, \pm 1,...$). The upper sign in Eqs.~(42) - (44) corresponds to
the FS domain  where $\mbox{sin}(\theta_2-\theta_1) = -1$ and the lower
sign corresponds to the FS$^{*}$ domain with
$\mbox{sin}(\theta_2-\theta_1) = 1$. This is a generalization of the
two-domain  FS phase discussed in Sec. III. The analysis of the
stability matrix (14) for these phase
 domains shows that FS is stable for $M > 0$ and FS$^{*}$ is stable for
$M<0$, just like our result in Sec. III. As these domains belong to the
same phase, namely, have the same free energy and are thermodynamically
equivalent, we shall consider one of them, for example, FS.

\subsection{Phase stability and phase diagram}

In order to outline the ($t,r$) phase diagram  we shall use the
 information given above for the other two phases which have their own
 domains of stability in the $(t,r)$ plane: N and FM. The
 FS stability conditions when $\gamma_1 \ne 0$ become

\begin{equation}
\label{eq45} 2 \gamma M -r -\gamma_1 M^2\ \ge 0,
\end{equation}
\begin{equation}
\label{eq46} \gamma M \ge 0,
\end{equation}
\begin{equation}
\label{eq47} 3(1-\gamma_1^2)M^2+3\gamma \gamma_1 M+t-\gamma_1
r-\gamma^2/2 \ge 0.
\end{equation}

and we prefer to  treat Eqs.~(45)~-~(47)~together with the
existence condition $\phi^2 \ge 0$, with $\phi$ given by Eq.~(42),
with the help of the picture shown in Fig.~4.

The most direct approach to analyze the existence and stability of FS
phase is to
 express $r$ as of function of $(M,t)$ from the equation of state (43),

\begin{eqnarray}
\label{48} \lefteqn{r^{(1)}_{eq}(t)=
}\\\nonumber&&\frac{M_{eq}}{(\gamma_1
M_{eq}-\gamma/2)}\left[(1-\gamma_1^2)M_{eq}^2+ \frac{3}{2} \gamma
\gamma_1 M_{eq} +(t-\frac{\gamma^2}{2})\right],
\end{eqnarray}

 and to substitute the above expression in the existence and
 stability conditions of FS-phase. It is obvious
 that there is a special value of $M$
\begin{equation}\label{49}
M_{S1}=\frac{\gamma}{2 \gamma_1}
\end{equation}
that is
 a solution of Eq.~(43) for any
 value of $r$ and
\begin{equation}\label{50}
  t_{S1}=-\frac{\gamma^2}{4\gamma_1^2},
\end{equation}
for which this procedure cannot be applied and should be considered
separately.  Note, that $M_{S1}$ is given by the respective horizontal
dashed line in Fig.~4.
 The analysis shows that in the interval $t_B^{(1)} < t<\gamma^2/2$ the phase
transition   is again of first order; here
\begin{equation}\label{51}
t_B^{(1)} =-\frac{\gamma^2}{4(1+\gamma_1)^2}.
\end{equation}
To find the equilibrium magnetization of first order phase transition,
depicted by the
  thick line $ACB$
 in Fig.~4 we need  the expression for equilibrium
free energy of FS-phase. It is obtained  from Eq.~(12) by setting
$(w=0,v=0)$ and substituting $r,\;\phi_i$ as given by Eqs.~(41),
(42) and (48). The result is
\begin{widetext}
\begin{equation}
\label{eq52} f^{(1)}_{FS} =
-\frac{M^2}{2(M\gamma_1-\gamma/2)^2}\times\left\{(1-\gamma_1^2)M^4 +
\gamma\gamma_1 M^3 + 2\left[t(1-\gamma_1^2)- \frac{\gamma^2}{8}\right]
M^2 - 2\gamma\gamma_1t M + t(t-\frac{\gamma^2}{2})\right\},
 \end{equation}
 \end{widetext}
where $M \equiv M_{eq}$.

For the phase transition from N to FS phase ($0<t<\gamma^2/2$),
$M_{eq}$  is found by setting the FS  free energy  from the above
expression equal to zero, as we have  by convention that  the free
energy of the normal phase is  zero. The value of $M^{(1)}_{eq}$
for positive $t$ is obtained numerically and is illustrated by
thick black curve $AC$ in Fig.~4.
 When
 $t^{(1)}_B\le t <0$ the transition is between FM and FS phases
 and we obtain $M^{(1)}_{eq}$ from the equation
 $f_{FS}=f_{FM}=(-t^2/2)$, where $f_{FM}$ is the free energy of FM
 phase. The equilibrium magnetization in the above $t$-interval is
 given by the formula
 \begin{equation}
\label{eq53} M_{eq}^{(1) \ast}=\frac{\gamma}{2(1+\gamma_1)},
\end{equation}
and  is drawn by thick line $CB$ in Fig. 4.

The existence and stability analysis shows that for
 $r>0$ the equilibrium
magnetization of the first order phase transition  should satisfy the
condition $ M^{(1)}_m < M^{(1)}_{eq}<M^{(1)}_0$.

By $M^{(1)}_0$ we denote  the positive solution of
$r^{(1)}(M_{eq})=0$ and its $t$-dependence is drawn in Fig.~4 by
the curve with circles. $M^{(1)}_m$ is the smaller positive root
of stability condition~(47) and also gives the maximum of function
$r^{(1)}_{eq}(M)$; see Eq.~(48). The  function $M^{(1)}_m$ is
depicted by the dotted curve  $AB$ in Fig.~4. When $t_{S1}<
t<t^{(1)}_B$ the existence and stability conditions are fulfilled
if $\sqrt{-t}<M<M_{S1}$, where $\sqrt{-t}$ is the magnetization of
ferromagnetic phase and is drawn by a thin black line on the left
of point B in Fig.~(4). Here we have two possibilities: $r>0$ for
$\sqrt{-t}<M <M^{(1)}_0$ and $r < 0$ for $M^{(1)}_0<M<M_{S1}$. To
the left of $t_{S1}$ and $t>t_{S2}$, where
\begin{equation}\label{54}
  t_{S2}= - \left(\frac{\gamma}{\gamma_1}\right)^2.
\end{equation}
the FS phase is stable and exists
 for $M_{S1}<M<\sqrt{-t}$. Here $r$ will be positive when
$M^{(1)}_0<M<\sqrt{-t}$ and  $r<0$ for $M^{(1)}_0>M>M_{S1}$. When
$t<t_{S2}$, $M<\sqrt{-t}$ and $r$ is always negative.

\begin{figure*}
\includegraphics{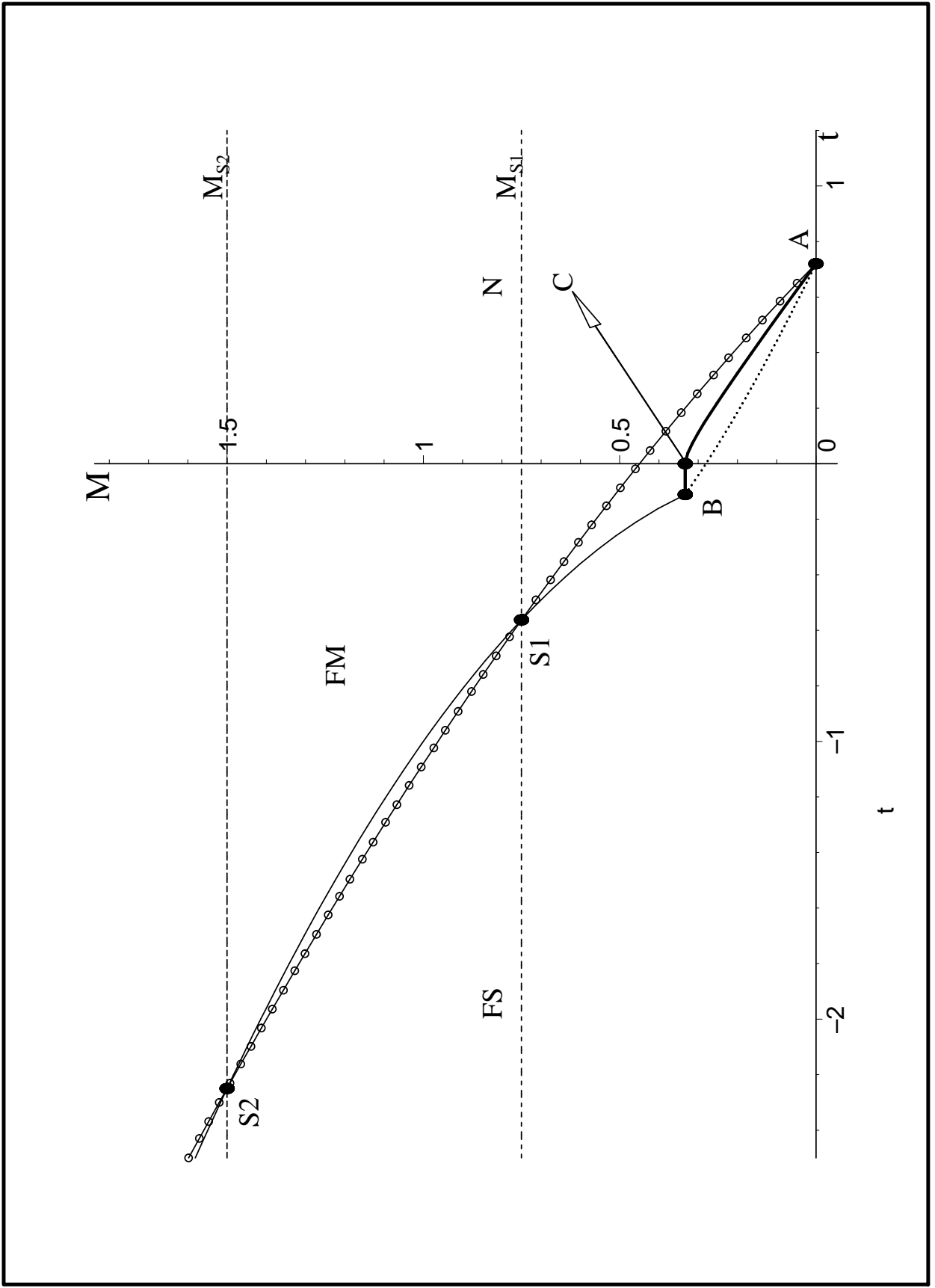}
\caption{\label{fig:wide}The dependence $M(t)$ as an illustration of
stability analysis for $\gamma=1.2$, $\gamma_1=0.8$ and $w=0$. The
parameters of the theory ($r,t,\gamma$, $\gamma_1$, $w,\dots$) are
defined by Eq.~(11). The horizontal dashed lines represent the
quantities $M_{\scriptsize S1}$ given by Eq.~(49) and $M_{\scriptsize
S2} = 2M_{\scriptsize S1}$. The line of circles $AS_1S_2$ describes the
positive solution of Eq.~(48). The thick line $AC$ gives the
equilibrium magnetization for $t>0$. The thick line $BC$ represents the
equilibrium magnetization for $t<0$ as given by Eq.~(53). The dotted
curve is the smaller positive solution of the stability condition (47).
The thin solid line $BS_1S_2$ is the magnetization $M=\sqrt{-t}$. The
arrow indicates the triple point $C$. $A$ and $B$ are tricritical
points of phase transition. The point $S_1$ corresponds to the maximum
of the curve (40) for $t < 0$, and the point $S_2$ corresponds to
$r_e^{(1)}(t) = 0$ in Eq.~(40).}
\end{figure*}

\begin{figure*}
\includegraphics{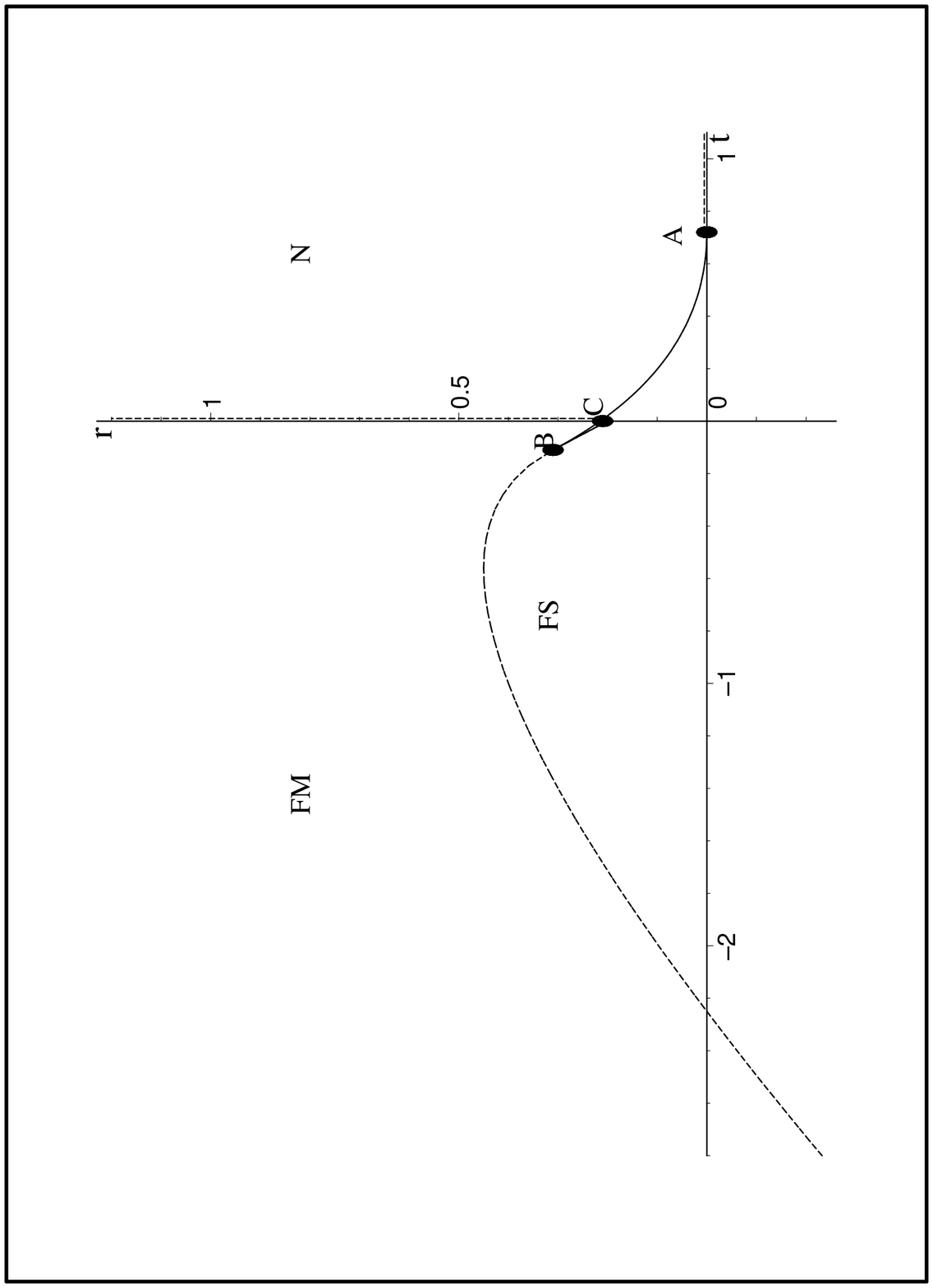}
\caption{\label{fig:wide}The phase diagram in the $(t,r)$ plane for
$\gamma=1.2,\;\gamma_1=0.8$ and $w=0$. The parameters of the theory
($r$, $t$, $\gamma$, $\gamma_1$, $w,\dots$) are defined by Eq.~(11).
The domains of stability of the phases N, FM and FS are indicated. $A$
and $B$ are tricritical points of phase transitions separating the
dashed lines (on the left of point $B$ and on the right of point $A$)
of second order phase transitions from the solid line $ABC$ of first
order phase transitions. The FS phase is stable in the whole domain of
the ($t$, $r$) below the solid and dashed lines. The vertical dashed
line coinciding with the $r$-axis above the triple point $C$ indicates
the N-FM phase transition of second order.}
\end{figure*}

On the basis of  the existence and stability analysis
 we draw in Fig.~5 the $(t,r)$-phase diagram  for concrete
values of $\gamma$ and $\gamma_1$. As we have mentioned above the order
of phase transitions is the same as for $\gamma_1=0$, see Fig.~3,
Sec.~III.
 The phase transition between the normal and FS phases is of
first order and goes along the equilibrium line $AC$ in the interval
($t_A=\gamma^2/2 $ and $t_C=0$). The function $r^{(1)}_{eq}(t)$ is
given by Eq. (48) with $M^{(1)}_{eq}$ from Fig. 4.

N, FM, and FS phases coexist at the triple point $C$ with coordinates
$t=0$, and $r_C^{(1)}=\gamma^2/4(\gamma_1+1)$. On the left of $C$ for
$t^{(1)}_B<t<0$ the phase transition line of first order
 $r^{(1)\ast}_{eq}(t)$ is found by substituting in Eq.~(48) the
 respective equilibrium magnetization, given by Eq. (53). In result we
 obtain
\begin{equation}
\label{eq55}
  r^{(1)\ast}_{eq}(t) = \frac{\gamma^2}{4(1+\gamma_1)}-t.
\end{equation}
This function is illustrated by the line $BC$ in Fig.~5 that terminates
at the tricritical point $B$ with coordinates $ t^{(1)}_B $ from
Eq.~(51), and
\begin{equation}\label{56}
  r^{(1)}_B=\frac{
\gamma^2(2+\gamma_1)}{4(1+\gamma_1)^2}.
\end{equation}
To the left of the tricritical point $B$ the second order phase
transition curve is given by the relation~(40). Here
 the magnetization is $M=\sqrt{-t}$ and the
superconducting order parameter is equal to zero
 ($\phi=0$). This line
intersects t-axis at $t_{S2}$ and is well defined also for $r<0$. The
function $r^{(1)}_{e}(t)$ has a maximum at the point $(t_{S1},
\gamma^2/4\gamma_1)$; here $M=M_{S1}$. When this point is approached
the second derivative of the free energy with respect to $M$ tends to
infinity. The result for the curves $r^{(1)}_{eq}(t)$ of equilibrium
phase transitions (N-FS and FM-FS) can be used to define the respective
equilibrium phase transition temperatures $T_{FS}$.

We shall not discuss the region, $t>0$, $r<0$, because we have
supposed from the very beginning that the transition temperature
for the ferromagnetic ordering T$_f$ is higher then the
superconducting transition temperature T$_s$, as is for the known
unconventional ferromagnetic superconductors. But this case may
become of substantial interest when, as one may expect, materials
with T$_f < $T$_s$ may be discovered experimentally.

\subsection{Discussion}

The shape of the equilibrium phase transition lines corresponding to
the phase transitions N-SC, N-FS, and FM-FS is similar to that of the
more simple case $\gamma_1 = 0$ and we shall not dwell on the variation
of the size of the phase domains with the variations of the parameter
$\gamma_1$ from zero to values constrained by the condition $\gamma_1^2
<1$. Our treatment from Sec. III of the magnetic susceptibility tensor
and the thermal quantities can be generalized in order to demonstrate
the dependence of these quantities on $\gamma_1$. We shall not consider
such problems. But an important qualitative difference between the
equilibrium phase transition lines shown in Figs.~3 and 5 cannot be
omitted. The second order phase transition line $r_e(t)$, shown by the
dotted line on the left of point $B$ in Fig.~3, tends to large positive
values of $r$ for large negative values of $t$ and remains in the
second quadrant ($t<0, r>0)$ of the plane ($t,r$) while the respective
second order phase transition line $r^{(1)}_{e}(t)$ in Fig.~5 crosses
the $t$-axis at the point $t_{S2}$ and is located in the third quadrant
($t<0,r<0$) for all possible values $t < t_{S2}$. This means that the
ground state (at 0 K) of systems with $\gamma_1 =0$ will be always the
FS phase whereas two types of ground states, FM and FS,
 can exist for systems with $0< \gamma_1^2 < 1$. The latter
seems more realistic when we compare theory and experiment, especially,
in ferromagnetic compounds like UGe$_2$, URhGe, and ZrZn$_2$.
Neglecting the $\gamma_1$-term does not allow to describe the
experimentally observed presence of FM phase at very low temperatures
and relatively low pressure $P$.

The final aim of the phase diagram investigation is the outline of the
($T,P$) diagram. Important conclusions about the shape of the $(T,P)$
diagram can be made from the form of the $(t,r)$ diagram without an
additional information about the values of the relevant material
parameters $(a_s$, $a_f,...$) and their dependence on the pressure $P$.
One should know also the characteristic temperature $T_s$, which has a
lower value than the experimentally observed~\cite{Saxena:2000,
Huxley:2001, Tateiwa:2001, Pfleiderer:2001, Aoki:2001} phase transition
temperature $(T_{FS} \sim 1 K)$ to the coexistence FS--phase. A
supposition about the dependence of the parameters $a_s$ and $a_f$ on
the pressure $P$ was made in Ref.~\cite{Walker:2002}. Our results for
$T_f \gg T_s$ show that the phase transition temperature $T_{FS}$
varies with the variation of the system parameters $(\alpha_s,
\alpha_f,...)$ from values which are  higher than the characteristic
temperature $T_s$ down to zero temperature. This is seen from Fig.~5.

In systems where a pure superconducting phase is not observed for
temperatures $T \sim T_f$ or $T\sim T_{\scriptsize FS}$, we can set
$T_s \sim 0$ in Eq. (21). Neglecting $T_s$ in Eq.~(21) and  assuming
that $(T^{\ast}/T_f) \ll 1$  we obtain that $T_c \equiv T_{\scriptsize
FS} \sim (T^{\ast}T_f)^{1/2}$. Note that the first
$(T^{\ast}/T_f)^{1/2}$-correction to this result has a negative sign
which means that a suitable dependence of the characteristic
temperature $T^{\ast}$ on the pressure P may be used in attempts to
describe the experimental shape of the FM-FS phase transition line in
the $(T,P)$ diagrams of UGe$_2$ and ZrZn$_2$; see, for example, Fig. 2
in Ref.~\cite{Saxena:2000}, Fig. 3 in Ref.~\cite{Huxley:2001}, Fig. 4
in Ref.~\cite{Pfleiderer:2001}. The experimental phase diagrams
indicate that $T_f(P)$ is a smooth monotonically decreasing function of
the pressure $P$ and $T_f(P)$ tends to zero when the pressure $P$
exceeds some critical value $P_c \sim 1$ GPa. Postulating the
respective experimental shape of the function $T_f(P)$ one may try to
give a theoretical prediction for the shape of the curve
$T_{\scriptsize FS}$ describing the FM-FS phase transition line. The
lack of experimental data about important parameters of the theory
forces us to make some suppositions about the behavior of the function
$T^{\ast}(P)$. The phase transition temperature $T_{\scriptsize FS}$
will qualitatively follow the shape of $T_f(P)$ provided the dependence
$T^{\ast}(P)$ is very smooth. This is in accord with the experimental
shapes of these curves near the critical pressure $P_c$ where both
$T_f$ and $T_{\scriptsize FS}$ are very small. The substantial
difference between $T_f$ and $T_{\scriptsize FS}$ at lower pressure ($P
< P_c$) can be explained with the negative sign of the correction term
to the leading dependence $T_{\scriptsize FS}(P) \sim
[T^{\ast}(P)T_f(P)]^{1/2}$ mentioned above and a convenient supposition
for the form of the function $T^{\ast}(P)$.

Eq.~(21) presents a rather simplified theoretical result for $T_C
\equiv T_{\scriptsize FS}$ because the effect of $M^2|\psi|^2$ coupling
is not taken into account. But following the same ideas, used in our
discussion of Eq.~(21), a more reliable theoretical prediction of the
shape of FM-FS phase transition line can be given on the basis of
Eq.~(40). Using the knowledge about the experimentally found shape of
$T_f(P)$ and the definition of the parameters $r$ and $t$ by Eq.~(11)
we substitute $T=T_{\scriptsize FS}(P)$ in Eq.~(40). In doing this we
have applied the following approximations, namely, that $T_s \sim 0$
for any pressure $P$, $T_{\scriptsize FS}(P_c) \sim T_f(P_c) \sim 0$
and for substantially lower pressure ($P < P_c$),  $T_f (P) \gg
T_{\scriptsize FS}(P)$. Then near the critical pressure $P_c$, we
easily obtain the transition temperature $T_{\scriptsize FS} \sim 0$,
as should be. For substantially lower values of the pressure there
exists an experimental requirement $(T_{\scriptsize FS} - T_s) \ll (T_f
- T_{\scriptsize FS})$. Using the latter we establish the approximate
formula $(T_f - T_{\scriptsize FS}) =
\gamma^2b_f^{1/2}/\gamma_1^2\alpha_f $. The same formula for $(T_f
-T_{\scriptsize FS})$ can be obtained from the parameter
$t_{\scriptsize S2}(T_{\scriptsize FS})$ given by Eq.~(54). The
pressure dependence of the parameters included in this formula defines
two qualitatively different types of behavior of $T_{\scriptsize
FS}(P)$ at relatively low pressures ($P \ll P_c$): (a) $T_{\scriptsize
FS}(P) \sim 0$ below some (second) critical value of the pressure
($P_c^{\prime} < P_c$), and (b) finite $T_{\scriptsize FS}(P)$ up to $P
\sim 0$. Therefore, we can estimate the value of the pressure
$P_c^{\prime} < P_c$ in UGe$_2$, where $T_{\scriptsize
FS}(P_c^{\prime}) \sim 0$. It can be obtained from the equation
$T_f(P_c^{\prime}) = (\gamma^2b_f^{1/2}/\gamma_1^2\alpha_f)$ provided
the pressure dependence of the respective material parameters is known.
So, the above consideration is consistent with the theoretical
prediction that the dashed line in Fig.~5 crosses the axis $r=0$ and
for this reason we have the opportunity to describe two ordered phases
at low temperatures and broad variations of the pressure. Our theory
allows also a description of the shape of the transition line
$T_{\scriptsize FS}(P)$ in ZrZn$_2$ and URhGe, where the transition
temperature $T_{\scriptsize FS}$ is finite at ambient pressure. To
avoid a misunderstanding, let us note that the diagram in Fig.~5 is
quite general and the domain containing the point $r=0$ of the phase
transition line for negative $t$ may not be permitted in some
ferromagnetic compounds.

Up to now we have discussed experimental curves of second order phase
transitions. Our analysis gives the opportunity to describe also first
order phase transition lines. Our investigation of the free energy (12)
leads to the prediction of triple ($C$)and tricritical points ($A$ and
$B$); see Figs. 3 and 5. We shall not dwell on the possible application
of these results to the phase diagrams of real substances, where first
order phase transitions and multicritical phenomena occur; see, e.g.,
Refs.~\cite{Kotegawa:2004, Huxley:2003}, where first order phase
transitions and tricritical points have been observed. The
consideration of such problems, in particular, the explanation of the
phase transition lines in Refs.~\cite{Kotegawa:2004, Huxley:2003}
requires further theoretical studies, that can be done on the basis of
a convenient extension of the free energy (12). For example, the
investigation of vortex phases in Ref.~\cite{Huxley:2003} needs taking
into account the gradient terms (4). Another generalization should be
done in order to explain the observation of two FM
phases~\cite{Kotegawa:2004, Huxley:2003}. Note, that the
experimentalists are not completely certain whether the FS phase is a
uniform or a vortex phase, and this is a crucial point for the
orientation of the further investigations. But we find quite
encouraging that our studies naturally lead to the prediction of the
same variety of phase transition lines and multicritial points that has
been observed in recent experiments~\cite{Kotegawa:2004, Huxley:2003}.

\section{ANISOTROPY EFFECTS}

Our analysis demonstrates that when the anisotropy of  Cooper
pairs is taken in consideration, there will be no drastic changes
in the shape the phase diagram for $r>0$ and the order of the
respective phase transitions. Of course, there will be some
changes in the size of the phase domains and the formulae for the
thermodynamic quantities. It is readily seen from Figs.~6 and 7
that the temperature domain of first order phase transitions and
the temperature domain of stability of FS above $T_s$ essentially
vary with the variations of the anisotropy parameter $w$. The
parameter $w$ will also insert changes in the values of the
thermodynamic quantities like the magnetic susceptibility and the
entropy and specific heat jumps at the phase transition points.

Besides, and this seems to be the main anisotropy effect, the $w$- and
$v$-terms in the free energy lead to a stabilization of the order along
the main crystal directions which, in other words, means that the
degeneration of the possible ground states (FM, SC, and FS) is
considerably reduced. This means also a smaller number of marginally
stable states.

\begin{figure*}
\includegraphics{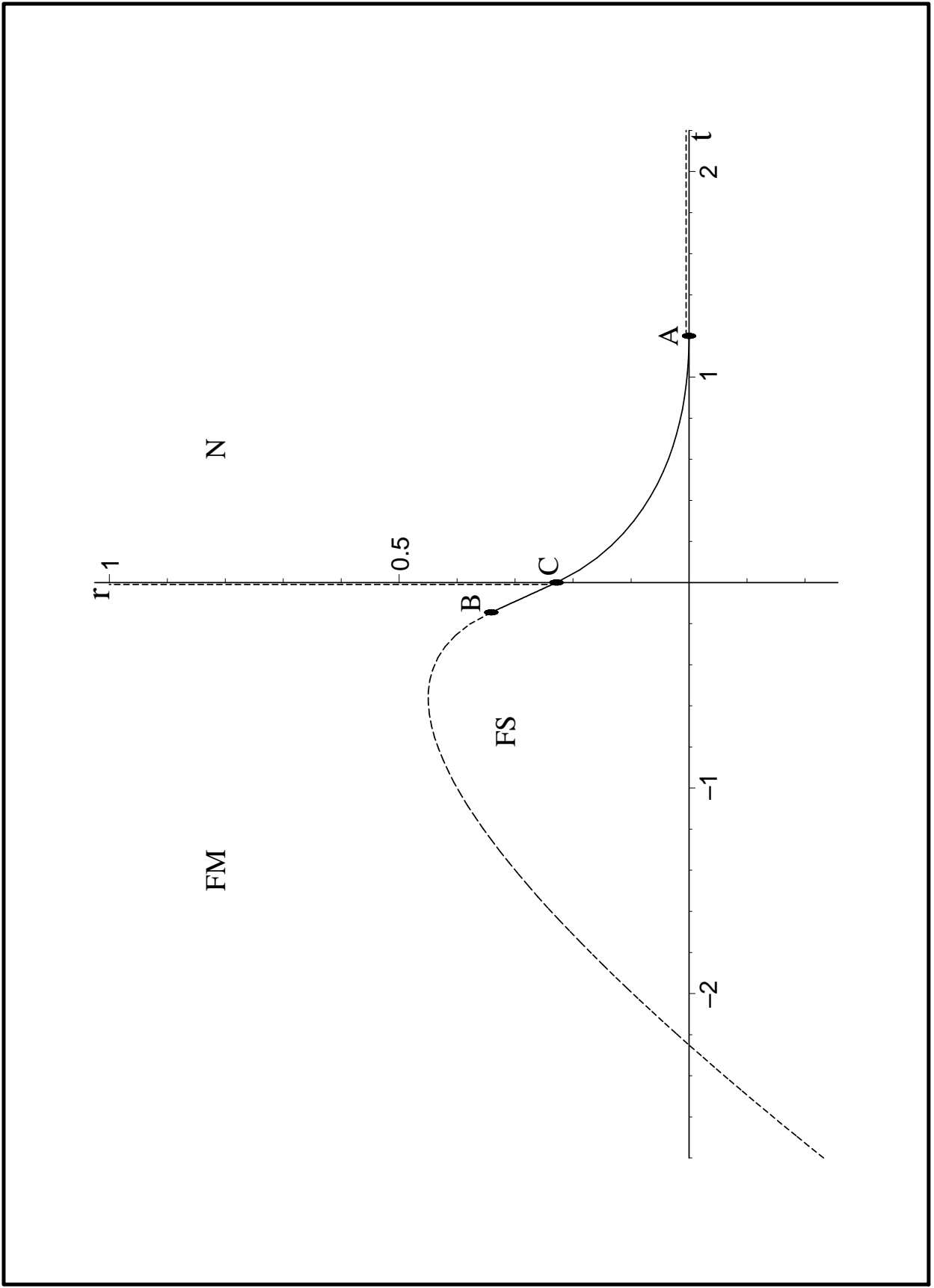}
\caption{\label{fig:wide}Phase diagram in the ($t$, $r$) plane for
$\gamma = 1.2, $ $\gamma_1 = 0.8$, and $w = 0.4$. The meaning of lines
and points is the same as given in Fig.~5.}
\end{figure*}

\begin{figure*}
\includegraphics{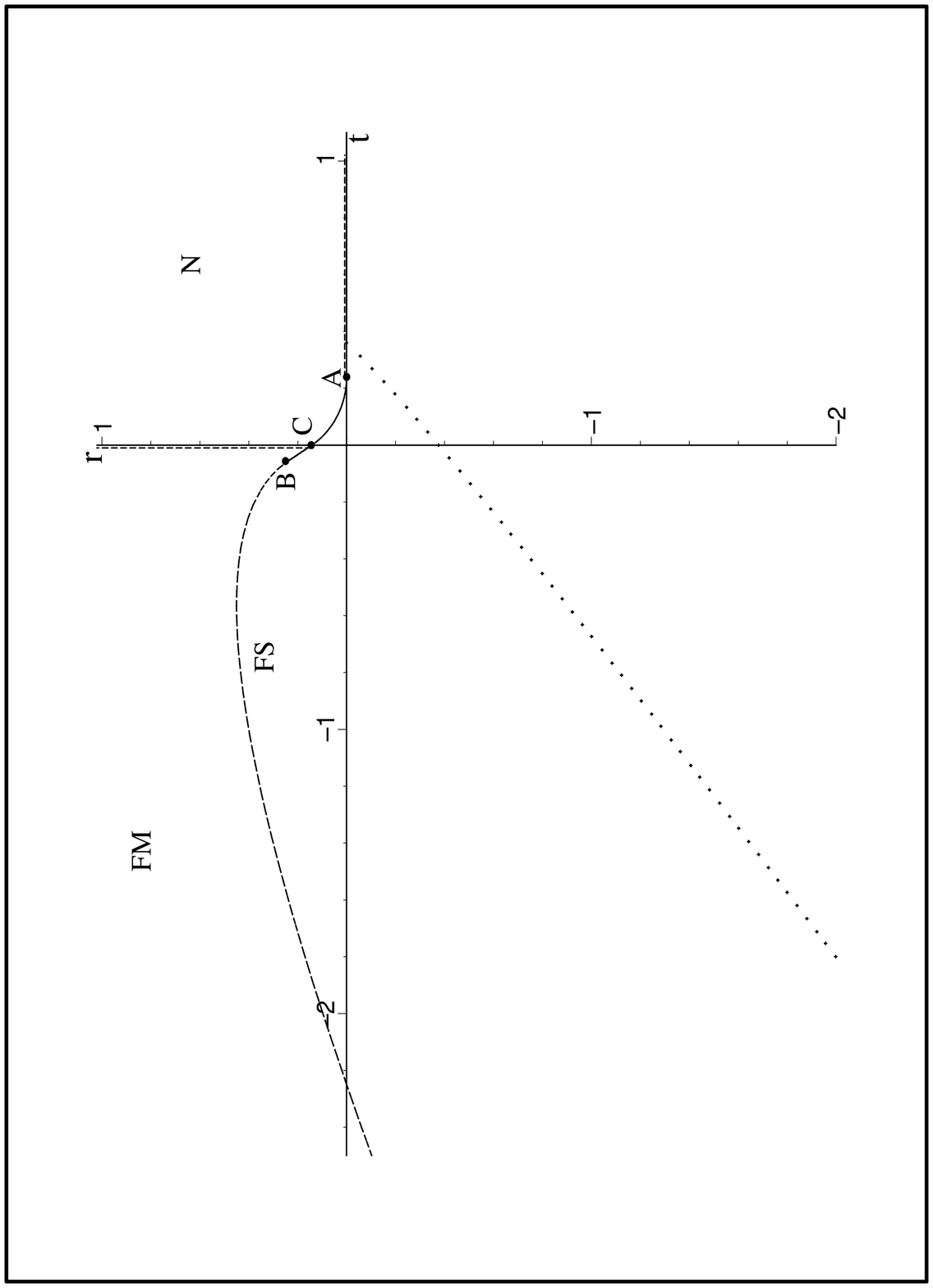}
\caption{\label{fig:wide}Phase diagram in the ($t$, $r$) plane for
$\gamma = 1.2, $ $\gamma_1 = 0.8$, and $w = -2$. The straight dotted
line for $r < 0$ indicates an instability of the FS phase. The meaning
of other lines and notations is the same as given in Fig.~5.}
\end{figure*}

The dimensionless anisotropy parameter $w=u_s/(b_s + u_s)$ can be
either positive or negative depending on the sign of $ u_s$. Obviously
when $ u_s > 0$, the parameter $w$ will be positive too and will be in
the interval $0<$ w$<1$ to ensure the positiveness of parameter $b$
from Eq.~(10).  When $w<0$, the latter condition is obeyed if the
original parameters of free energy (3) satisfy the inequality
$-b_s<u_s<0$.

We should mention here that a new phase of coexistence of
superconductivity and ferromagnetism occurs as a solution of Eqs.~(13).
It is defined in the following way:
\begin{eqnarray}
\label{eq57} \phi^2_1+\phi^2_2 & = &
\frac{1}{1-\gamma_1^2}\left[\gamma_1(t+\frac{\gamma^2}{2w})
-r\right],\\ \nonumber M^2&= &
\frac{1}{1-\gamma_1^2}\left[\gamma_1r-(t+\frac{\gamma^2}{2w}) \right],
\end{eqnarray}
and
\begin{equation}
\label{eq58} 2w\sin{(\theta_2-\theta_1)}=\gamma M\:,\;\;\;
\cos{(\theta_2-\theta_1)}\ne 0 \:.
\end{equation}
 In the present
approximation the phase~(57)-(58) is unstable, but this may be changed
when crystal anisotropy is taken into account.

	We shall write
  the equations for order parameters $M$ and
  $\phi_j$ of FS phase  in order to illustrate the changes when $w\ne0$
 \begin{equation}
\label{eq59} \phi^2=\frac{\pm \gamma M-r-\gamma_1 M^2}{(1-w)} \ge 0,
\end{equation}

and
\begin{eqnarray}
\label{eq60} \lefteqn{(1- w - \gamma_1^2)M^3}\\\nonumber&&\pm
\frac{3}{2} \gamma \gamma_1 M^2
+\left[t(1-w)-\frac{\gamma^2}{2}-\gamma_1 r\right]M\pm \frac{\gamma
r}{2}=0,
\end{eqnarray}

where the meaning of the upper and lower sign is
the same as explained just below Eq.~(44).  The difference in the
stability conditions is more pronounced and gives new effects that
will be explained further,
\begin{equation}
\label{eq61} \frac{ (2-w)\gamma M- r -\gamma_1M^2}{1-w} \ge 0,
\end{equation}
\begin{equation}
\label{eq62} \gamma M -wr-w \gamma_1 M^2 \geq 0,
\end{equation}
and

\begin{equation}
\label{eq63} \frac{3(1-w-\gamma_1^2) M^2 + 3 \gamma \gamma_1 M +
t(1-w)-\gamma^2/2 -\gamma_1 r}{1-w}\geq 0.
\end{equation}

The calculations of the phase diagram in ($t,\;r$) parameter space are
done in the same way as in case of  $w=0$ and show that for $w>0$ there
is no qualitative change of the phase diagram. Quantitatively, the
region of first order phase transition widens both with respect to $t$
and $r$ as illustrated in Fig.~6. On the contrary, when $w < 0$ the
first order phase transition region becomes more narrow but the
condition~(62) limits the stability of FS  for $r<0$. This is seen from
Fig.~7 where FS is stable above the straight dotted line for $r < 0$
and $t < 0$. So, purely superconducting (Meissner) phases occur also as
ground states together with FS and FM phases.

\section{CONCLUSION}

We investigated  the M-trigger effect in unconventional ferromagnetic
superconductors. This effect arises from  the $M\psi_1\psi_2$-coupling
term in the GL free energy and brings into existence a
superconductivity in a domain of the system's phase diagram   that is
entirely occupied by the ferromagnetic phase. The coexistence of
unconventional superconductivity and ferromagnetic order is possible
for temperatures above and below the critical temperature $T_s$, that
corresponds to the standard second-order phase transition
 from normal to Meissner phase -- usual uniform superconductivity
in a zero external magnetic field which occurs outside the domain of
existence of ferromagnetic order. Our investigation has been mainly
intended to clarify the thermodynamic behavior at temperatures $T_s< T
< T_f$ where the superconductivity cannot appear without the mechanism
of M-triggering. We have described the possible ordered phases (FM and
FS) in this most interesting temperature interval.

The Cooper pair and crystal anisotropies have also been investigated
and their main effects on the thermodynamics of the triggered phase of
coexistence is established. In discussions of concrete real material
one should consider the respective crystal symmetry. But when the low
symmetry and low order (in both  $M$ and $\psi$) $\gamma$-term is
present in the free energy, the dependence of essential thermodynamic
properties on the type of crystal symmetry is not substantial.

Below the superconducting critical temperature $T_s$ a variety of pure
superconducting and mixed phases of coexistence of superconductivity
and ferromagnetism exists and the thermodynamic behavior at these
relatively low temperatures is more complex than in known cases of
improper ferroelectrics. The case $T_f < T_s$ also needs a special
investigation.

Our results are referred to the possible uniform superconducting and
ferromagnetic states. Vortex and other nonuniform phases need a
separate study.

The relation of the present investigation to properties of real
ferromagnetic compounds, such as UGe$_2$, URhGe, and ZrZn$_2$, has been
discussed throughout the text. In these compounds the ferromagnetic
critical temperature is much larger than the superconducting critical
temperature $(T_f \gg T_s)$ and that is why the M-triggering of the
spin-triplet superconductivity is very strong. Moreover, the
$\gamma_1$-term is important to stabilize the FM order up to the
absolute zero (0 K), as is in the known spin-triplet ferromagnetic
superconductors. Ignoring~\cite{Walker:2002}  the symmetry conserving
$\gamma_1$-term does not allow a proper description of the known real
substances of this type. More experimental information about the values
of the material parameters ($a_s, a_f, ...$) included in the free
energy (12) is required in order to outline the thermodynamic behavior
and the phase diagram in terms of thermodynamic parameters $T$ and $P$.
In particular, a reliable knowledge about the dependence of the
parameters $a_s$ and $a_f$ on the pressure $P$, the value of the
characteristic temperature $T_s$ and the ratio $a_s/a_f$ at zero
temperature are of primary interest. \\

{\bf ACKNOWLEDGEMENTS}\\

One of us (D.I.U.) thanks the  hospitality of ICTP (Triest) and MPI-PKS
(Dresden), where a part of this work has been made. Financial support
by SCENET (Parma)  is also acknowledged.\\

\end{document}